\RequirePackage{arydshln}
\documentclass[aps,twocolumn,nofootinbib,superscriptaddress,preprintnumbers,pra,10pt,floatfix]{revtex4-1}

\usepackage[utf8]{inputenc}

\usepackage{float}
\usepackage{amsmath,amssymb}
\usepackage{dsfont} 
\usepackage{hyperref}
\usepackage{graphicx}
\usepackage{enumitem}
\usepackage{mathtools}
\usepackage{bbold}
\usepackage{multirow}
\usepackage{ytableau}
\usepackage{youngtab}
\usepackage{braket}
\usepackage{soul}
\usepackage{cancel}
\usepackage{xcolor}
\usepackage[normalem]{ulem}

\usepackage{lipsum}

\usepackage{colortbl} 
\definecolor{Gray}{gray}{0.95}
\definecolor{RGray}{gray}{0.90}
\definecolor{CGray}{gray}{0.92}

\usepackage{arydshln}

\makeatletter
\g@addto@macro\bfseries{\boldmath}
\makeatother

\makeatletter
\renewcommand\paragraph{\@startsection{paragraph}{4}{\z@}%
                                    {3.25ex \@plus1ex \@minus.2ex}%
                                    {-1em}%
                                    {\normalfont\normalsize\bfseries}}
\makeatother

\allowdisplaybreaks
\begin{document}

\preprint{}
\preprint{}

\title{ Revisiting  $B\to K^{(\ast)} \nu \bar{\nu}$ decays in the Standard Model and beyond}

\author{Damir Be\v{c}irevi\'c}
\email{damir.becirevic@ijclab.in2p3.fr}
\affiliation{IJCLab, P\^ole Th\'eorie (Bat.~210), CNRS/IN2P3 et Universit\'e, Paris-Saclay, 91405 Orsay, France}
\author{Gioacchino Piazza}
\email{gioacchino.piazza@ijclab.in2p3.fr}
\affiliation{IJCLab, P\^ole Th\'eorie (Bat.~210), CNRS/IN2P3 et Universit\'e, Paris-Saclay, 91405 Orsay, France}
\author{Olcyr Sumensari}
\email{olcyr.sumensari@ijclab.in2p3.fr}
\affiliation{IJCLab, P\^ole Th\'eorie (Bat.~210), CNRS/IN2P3 et Universit\'e, Paris-Saclay, 91405 Orsay, France}

\begin{abstract}
\vspace{5mm}
In this letter we revisit the Standard Model predictions for $\mathcal{B}(B\to K^{(\ast)}\nu\bar{\nu})$ and discuss the opportunities that open up when combining its partial decay rate with that of $B\to K^{(\ast)}\ell\ell$. In the Standard Model a suitable ratio of these two modes can be used to extract $C_9^\mathrm{eff}$, which is essential for a reliable phenomenological analysis of the $B\to K^{(\ast)}\ell\ell$ angular observables. The same ratio also proves to be more sensitive to the presence of New Physics in many plausible extensions of the Standard Model. 
We also suggest that the separate measurement of $\mathcal{B}(B\to K\nu\bar{\nu})$ for high and for low $q^2$'s can be helpful for testing the assumed shape of the vector form factor, because the lattice QCD data are obtained at high $q^2$'s, whereas the low $q^2$ region is obtained through an extrapolation. 
\vspace{3mm}
\end{abstract}

\maketitle

\allowdisplaybreaks

\section{Introduction}\label{sec:intro}

Over the past decade a great effort in the high energy physics community has been invested in studying the exclusive decays based on the $b\to s ll$ mode, with $l\in \{e,\mu\}$. Mediated by the flavor changing neutral current, these decays can occur only through loops in the Standard Model (SM), and therefore their measurement was expected to give us insight into the loop content, which in turn could reveal a presence of physics beyond the SM (BSM). The LHC experiments performed detailed studies of $B\to K^\ast (\to K\pi) \mu\mu$~\cite{LHCb:2015svh,CMS:2020oqb} and $B\to K \mu\mu$~\cite{LHCb:2014auh,CMS:2018qih}. The angular distribution of these decays offered access to a set of new observables, free of the Cabibbo-Kobayashi-Maskawa (CKM) uncertainties, each with an additional and often complementary way to test the presence of the BSM physics~\cite{Becirevic:2011bp,Descotes-Genon:2013vna}. It soon became clear, however, that the main obstacles to such an endeavor are the hadronic uncertainties. On the one hand and despite the improvement in controlling the uncertainties of hadronic matrix elements of the local operators, an $\mathcal{O}(10\%)$ uncertainty can be warrantied only in a few cases. On the other hand, the matrix element of the non-local operators, arising from couplings to the $c\bar c$-pairs, remains an open problem, see e.g.~Ref.~\cite{Ciuchini:2015qxb}. To get around the latter problem one tries to stay below the region populated by the $c\bar c$-resonances. To evaluate the matrix element of the non-local operator one then opts to either invoke the quark-hadron duality which then means relying entirely on the perturbation theory~\cite{Greub:2008cy,Asatryan:2001zw,Asatrian:2019kbk}, or to employ a (hadronic) model calculation~\cite{Khodjamirian:2010vf,Khodjamirian:2012rm,Gubernari:2020eft,Gubernari:2022hxn}. The problems related to both kinds of hadronic uncertainties appear to be almost entirely absent in the measurement of the ratio $\smash{R_{K^{(\ast)}} = \mathcal{B}^\prime(B\to K^{(\ast )} \mu\mu)/ \mathcal{B}^\prime(B\to K^{(\ast )} e e )}$~\cite{Hiller:2003js}, where $\mathcal{B}^\prime$ is used to indicate that the partial branching fractions are measured in the interval $q^2\in [1.1, 6]~\mathrm{GeV}^2$, therefore below the first $c\bar c$ resonance, $m_{J/\psi}^2 = (3.097\,\mathrm{GeV})^2$. Indeed, the measurement of $R_K$ and $R_{K^\ast}$~\cite{LHCb:2022qnv}, and of $\mathcal{B}(B_s\to \mu\mu)$~\cite{ATLAS:2018cur,CMS:2020rox,LHCb:2021awg} resulted in clean constraints on the BSM couplings and on the models of physics BSM. Many BSM scenarios predict a significant deviation of $\mathcal{B}(B\to K^{(\ast )} \nu\bar\nu)$ with respect to its SM prediction and therefore $\mathcal{B}(B\to K^{(\ast )} \nu\bar\nu)$ can provide us with either a test of validity of a given model, or with a constraint when building an acceptable scenario of physics BSM.  

In this letter we will first evaluate $\mathcal{B}(B\to K^{(\ast )} \nu\bar\nu)$ in the SM by arguing that the most precise information can be obtained if one splits the $B\to K \nu\bar\nu$ events to those with high- and those with low-$q^2$'s. From the comparison of the measured $\mathcal{B}(B\to K  \nu\bar\nu)_{\mathrm{high}-q^2}$ with the one predicted in the SM, one can check for a consistency with the SM. Instead, from the ratio $\mathcal{B}(B\to K  \nu\bar\nu)_{\mathrm{low}-q^2}/\mathcal{B}(B\to K  \nu\bar\nu)_{\mathrm{high}-q^2}$ one can check on the validity of the shape of the sole form factor entering the expression for $\mathrm{d}\mathcal{B}(B\to K  \nu\bar\nu)/\mathrm{d}q^2 \propto | f_+(q^2)|^2$. 
We will then argue, like in Ref.~\cite{Bartsch:2009qp}, that the ratio $\mathcal{B}^\prime(B \to K\mu \mu)/\mathcal{B}^\prime(B \to K\nu \bar \nu)$, in the low $q^2$-bin, is essentially free of the form factor uncertainty and, if measured, it allows us to extract the desired Wilson coefficient $C_9^\mathrm{eff}$, which is the one plagued by uncertainties arising from the hadronic matrix element of the non-local operator.  
A similar discussion, even if somewhat less accurate, can then be extended to $B\to K^{\ast } \nu\bar\nu$. In that way one can also assess the size of the non-factorizable contribution to the $C_9^\mathrm{eff}$ which, according to Ref.~\cite{Khodjamirian:2010vf}, is different in the case of $K$ from the case of $K^\ast$ in the final state.   

On the basis of that information, the controversies in comparison of other observables extracted from the experimental angular analysis of $B\to K^{(\ast)} \mu\mu$ with their SM values would be removed, and the search of a scenario of physics BSM consistent with many more experimental constraints would become more compelling. We show that a study of the ratio $\mathcal{B}^\prime(B \to K\mu \mu)/\mathcal{B}^\prime(B \to K\nu \bar \nu)$ could provide us with a useful filter to select among the acceptable models of physics BSM.

The remainder of this letter is organized as follows: 
In Sec.~\ref{sec:BKnunu} we describe the $B\to K^{(\ast)} \nu \bar{\nu}$ decays in the SM, with a focus on hadronic uncertainties. In Sec.~\ref{sec:improved}, we propose alternative observables that are potentially less sensitive to hadronic uncertainties, and we discuss their sensitivity to physics beyond the SM in Sec.~\ref{sec:bsm}.
Our findings are briefly summarized in Sec.~\ref{sec:summary}.


\section{$B\to K^{(\ast)} \nu \bar{\nu}$ Decays in the SM\label{sec:BKnunu}}

\subsection{Effective theory  description}\label{ssec:SM-description}

Decays based on the $b\to s \nu \bar{\nu}$ transition are described by the following effective Lagrangian,
\begin{align}
\label{eq:eft-bsnunu}
\mathcal{L}_\mathrm{eff}^{\mathrm{b\to s\nu\nu}} =  \dfrac{4 G_F}{\sqrt{2}} \lambda_t \sum_a C_a\, \mathcal{O}_a+\mathrm{h.c.}\,, 
\end{align}

\noindent where $G_F$ is the Fermi constant, $\lambda_t = V_{tb} V_{ts}^\ast$ is a suitable product of the CKM entries, and the only relevant operator in the SM is given by~\footnote{The right-handed operator $\mathcal{O}_{R}^{\nu_i\nu_j} =\frac{e^2}{(4\pi)^2}(\bar{s}_R \gamma_\mu b_R)(\bar{\nu}_i \gamma^\mu (1-\gamma_5)\nu_j)$ is absent in the SM, but it can appear in some of the BSM scenarios (see Sec.~\ref{sec:bsm}).}
\begin{align}
\mathcal{O}_{L}^{\nu_i\nu_j} &=\dfrac{e^2}{(4\pi)^2}(\bar{s}_L \gamma_\mu b_L)(\bar{\nu}_i \gamma^\mu (1-\gamma_5)\nu_j)\,.
\end{align}

\noindent 
The SM effective Wilson coefficient $\big{[}C_{L}^{\nu_i\nu_j}\big{]}_\mathrm{SM}\equiv\delta_{ij}\, C_{L}^{\mathrm{SM}}$ is known~\cite{Buras:2014fpa},
\begin{align}
C_{L}^\mathrm{SM} = -X_t/\sin^2\theta_W\,, \qquad X_t=1.462(17)(2)\,,
\end{align}

\noindent and it includes the NLO QCD corrections~\cite{Buchalla:1993bv,Buchalla:1998ba,Misiak:1999yg}, as well as the two-loop electroweak contributions~\cite{Brod:2010hi}. Using $\sin^2\theta_W=0.23141(4)$~\cite{ParticleDataGroup:2022pth}, one finally arrives at $C_{L}^{\mathrm{SM}}=-6.32(7)$, where the dominant source of uncertainty comes from the higher order QCD corrections.

\paragraph*{\underline{$B\to K \nu\bar{\nu}$}} The SM differential decay rate of $B\to K\nu\bar{\nu}$ can be written as
\begin{align}
\label{eq:dBR}
\hspace{-0.15cm}\dfrac{\mathrm{d}\mathcal{B}}{\mathrm{d}q^2} (B\to K\nu\bar{\nu})= &  \mathcal{N}_K(q^2)\, |C_L^\mathrm{SM}|^2 \, |\lambda_t|^2 \left[f_+(q^2)\right]^2\,,
\end{align}
where $0 < q^2\leq (m_B-m_K)^2$ is the di-neutrino invariant mass, $f_+(q^2)$ is the $B\to K$ vector form factor which will be discussed in Sec.~\ref{ssec:form factors}, and $\mathcal{N}_K(q^2)$ denotes a known $q^2$-dependent function,
\begin{align}
\mathcal{N}_K(q^2) = \tau_{B} \dfrac{G_F^2\,\alpha_\mathrm{em}^2}{ 256 \pi^5}\dfrac{\lambda^{3/2}_K}{m_B^3}\,,
\end{align}

\noindent with  $\lambda_K\equiv\lambda(q^2,m_B^2,m_K^2)$ being the triangle function  $\lambda(a^2,b^2,c^2)\equiv\left(a^2-(b-c)^2\right)\left(a^2-(b+c)^2\right)$. Note that in the above expressions we summed over the neutrino flavors. 

\paragraph*{\underline{$B\to K^\ast \nu\bar{\nu}$}} Similarly to the previous case, the $B\to K^\ast \nu\bar{\nu}$ branching fraction can be written as:
\begin{align}
\label{eq:dBRst}
\hspace{-0.35cm}\dfrac{\mathrm{d}\mathcal{B}}{\mathrm{d}q^2} (B\to K^\ast\nu\bar{\nu})= &  \mathcal{N}_{K^\ast}(q^2) |C_L^\mathrm{SM}|^2  |\lambda_t|^2 \mathcal{F}(q^2)\,,
\end{align}
where the kinematical factor reads,
\begin{align}
\mathcal{N}_{K^\ast}(q^2) =\tau_{B} \dfrac{G_F^2\,\alpha_\mathrm{em}^2}{128 \pi^5}\dfrac{\lambda^{1/2}_{K^\ast} q^2}{m_B^3}\left(m_B + m_{K^\ast} \right)^2 \,,
\end{align}
with $\lambda_{K^\ast}\equiv\lambda(q^2,m_B^2,m_{K^\ast}^2)$, and $\mathcal{F}(q^2)$ given by
\begin{align}
\label{eq:FF}
\begin{split}
\mathcal{F}(q^2)=
[A_1(q^2)]^2&+ \dfrac{32\ m_{K^\ast}^2 m_B^2}{q^2 (m_B+m_{K^\ast})^2 }[A_{12}(q^2)]^2\\
&+\dfrac{\lambda_{K^\ast}}{(m_B+m_{K^\ast})^4}[V(q^2)]^2\,.
\end{split}
\end{align}
\noindent The $B\to K^\ast$ form factors $A_1(q^2)$, $A_{12}(q^2)$ and $V(q^2)$ will be defined shortly, in Sec.~\ref{ssec:form-factors-BKst}. 

\

Besides the small and controlled uncertainty in $C_{L}^{\mathrm{SM}}$, two other sources of theoretical uncertainties in the above expressions come from: (i) the $B\to K^{(\ast)}$ form factors that must be determined nonperturbatively, see Sec.~\ref{ssec:form factors} and~\ref{ssec:form-factors-BKst}, and (ii) the product of the CKM matrix elements $\lambda_t=V_{tb} V_{ts}^\ast$, which will be discussed in Sec.~\ref{ssec:ckm}. 

\subsection{$B\to K$ form factors}\label{ssec:form factors}

The $B\to K$ hadronic matrix element is the main source of theoretical uncertainty entering the $B\to K \nu\bar{\nu}$ branching fraction. It is usually decomposed as
\begin{align}\label{eq:ff}
\langle \bar{K}(k) | \bar{s}\gamma^\mu b | \bar{B}(p) \rangle &= \Big{[}(p+k)^\mu- \dfrac{m_B^2-m_K^2}{q^2}q^\mu\Big{]} f_{+}(q^2) \nonumber \\&+ \dfrac{m_B^2-m_K^2}{q^2} q^\mu f_0(q^2)\,,
\end{align}

\noindent where $f_+$ ($f_0$) are the so-called vector (scalar) $B\to K$ form factors, satisfying at $q^2=0$ the condition $f_+(0)=f_0(0)$. Note also that the scalar form factor does not enter the theoretical expression for $\mathcal{B}(B\to K \nu\bar{\nu})$, cf.  Eq.~\eqref{eq:dBR}, since the neutrino masses are negligible.

\begin{figure}[t!]
\centering
\includegraphics[width=.98\linewidth]{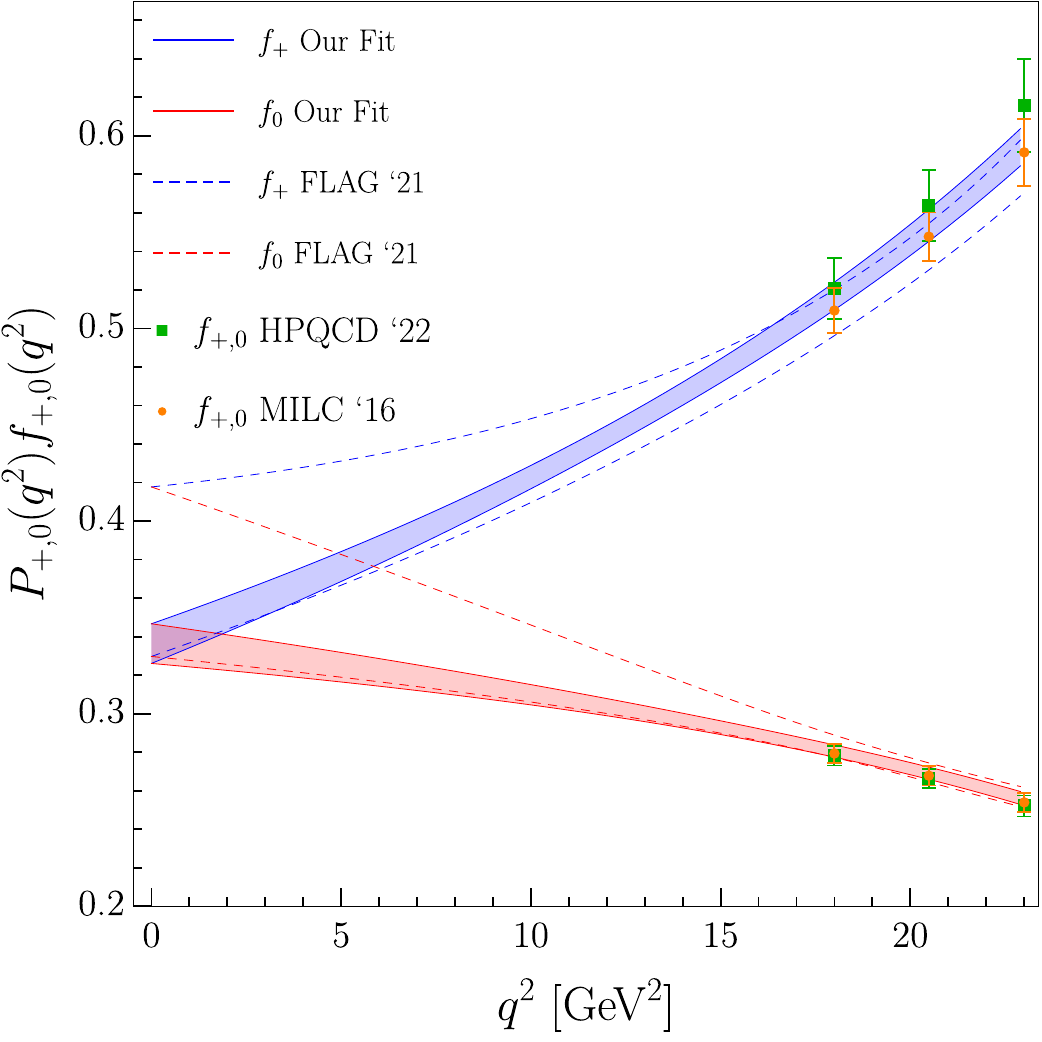}
\caption{\small \sl
The results of our fit for $f_{+}(q^2)$ and $f_{0}(q^2)$ form factors are depicted by the blue and red solid curves respectively. The dashed lines correspond to the results reported by FLAG~\cite{Aoki:2021kgd}. The synthetic data points by HPQCD (green)~\cite{Parrott:2022rgu} and by FNAL/MILC~(orange)~\cite{Bailey:2015dka} are also shown for comparison. $P_{+,0}(q^2)$ are the inverse pole terms defined in Eq.~\eqref{eq:poleP}.}
\label{fig:form factors} 
\vspace*{1.em}
\includegraphics[width=0.98\linewidth]{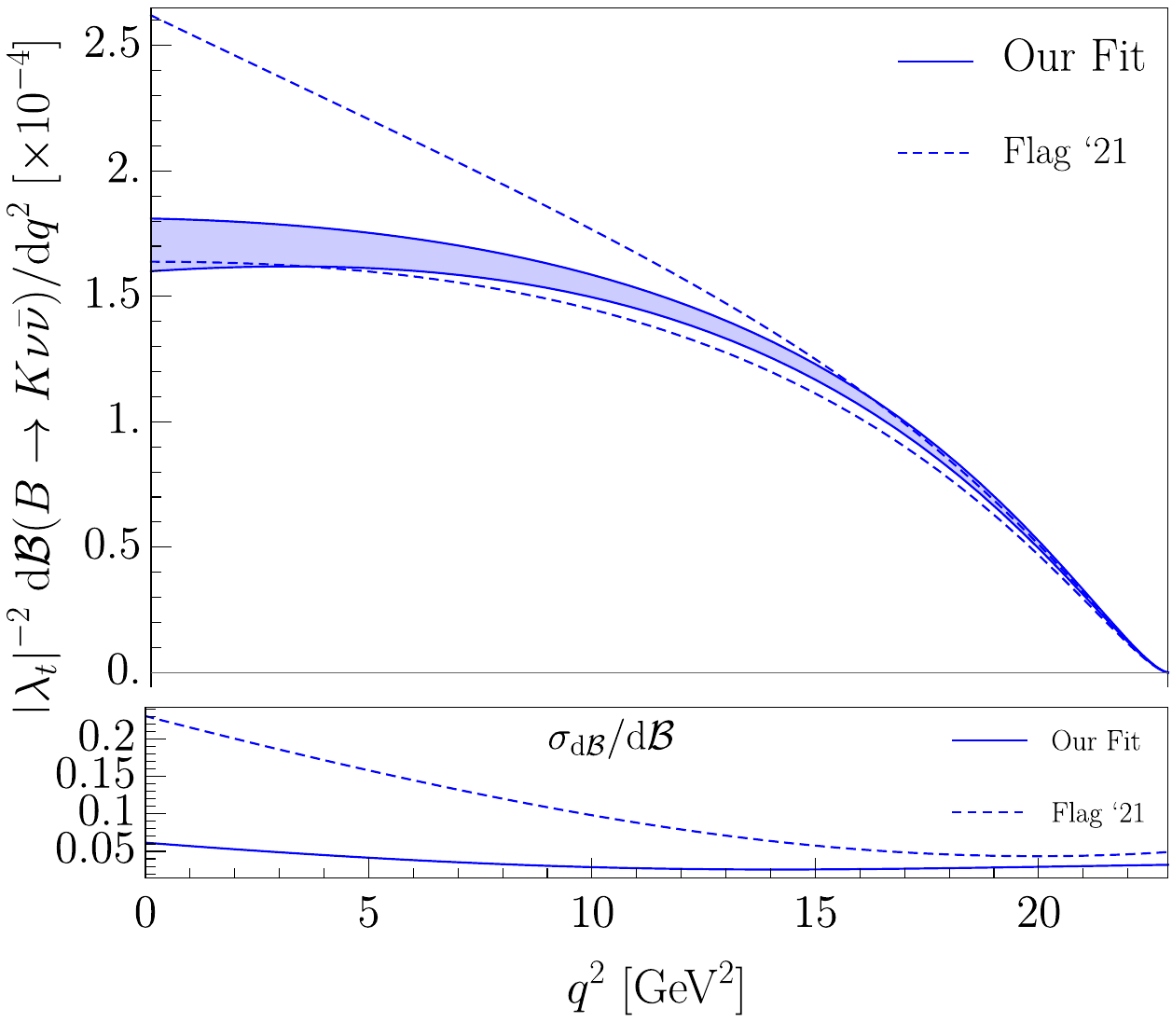}
\caption{\small \sl Rescaled differential branching fraction $|\lambda_t|^{-2}\,\mathrm{d}\mathcal{B}(B^+\to K^+ \nu\bar{\nu})/\mathrm{d}q^2$ is plotted by using the form factor $f_+(q^2)$ from the FLAG review~\cite{Aoki:2021kgd}, and by using the one discussed in this letter. The lower panel shows the relative uncertainty on this quantity as a function of $q^2$ using the FLAG form factors. }
\label{fig:dBR} 
\end{figure}

While this letter was in writing an update of the lattice QCD result by the HPQCD collaboration appeared~\cite{Parrott:2022rgu}. We used the information provided in their paper, combined it with the lattice QCD results by the FNAL/MILC collaboration~\cite{Bailey:2015dka} and followed the same procedure as  FLAG~\cite{Aoki:2021kgd} in order to provide the new average of the lattice QCD form factors, cf. Appendix~\ref{app:form factors}. Our average therefore supersedes the one presented in Ref.~\cite{Aoki:2021kgd}, in which now already obsolete HPQCD results from Ref.~\cite{Bouchard:2013eph} have been used. 
Since the lattice QCD results are obtained for $q^2 \gtrsim 16~\mathrm{GeV}^2$, an extrapolation is needed to cover the entire $B\to K\nu\bar{\nu}$ physical region. 
This is provided by the parametrization of the $q^2$ dependence of the form factors, which is discussed in Ref.~\cite{Aoki:2021kgd}.  
In Fig.~\ref{fig:form factors} we show the newly averaged form factors and their shapes (solid curves), and compare them with the previous ones (dashed) presented in~\cite{Aoki:2021kgd}. Clearly, the effect of the inclusion of the new HPQCD results is that the form factors at low $q^2$'s are now more accurate. 
One should keep in mind, however, that the low-$q^2$ region is a result of an extrapolation which is a potential source of systematic uncertainty, not accounted in the error budget. We propose a way to monitor such an uncertainty by splitting the sample of $B\to K \nu\bar{\nu}$ events into two bins: low $q^2$'s for $q^2/(m_B-m_K)^2 \in (0,1/2)$, and high $q^2$'s for $q^2/(m_B-m_K)^2 \in (1/2,1)$.~\footnote{Here we propose to split the sample in two bins, but obviously splitting the sample into more than two bins would be even better when it comes to monitoring the shape of the form factor.} From the measured ratio:
\begin{align}
\label{eq:rlh}
r_\mathrm{lh}={\mathcal{B}(B \to K  \nu\bar{\nu})_{\mathrm{low}-q^2}
\over \mathcal{B}(B\to K \nu\bar{\nu})_{\mathrm{high}-q^2}},
\end{align}
one can check whether or not the result is consistent with prediction, which is obtained by using lattice QCD results for $f_+(q^2)$ at high $q^2$, and the ones obtained through extrapolation to low $q^2$. Note that the above ratio $r_\mathrm{lh}$ is independent on the CKM factor, and on the Wilson coefficient regardless of the presence of physics BSM, provided we consider the left-handed neutrinos. Using the new average for the form factor provided in this paper we find, 
\begin{align}
r_\mathrm{lh} = 1.91(6)\,,
\end{align}
which is obviously consistent with $r_\mathrm{lh}=1.92(20)$, obtained by using solely the FNAL/MILC form factors~\cite{Bailey:2015dka}, as well as with $r_\mathrm{lh}=1.85(9)$, that we obtain by using only the new HPQCD values~\cite{Parrott:2022rgu}. 
Finally, and for future reference we note that from our average $f_+(0) = 0.336(10)$, which again is consistent with the FNAL/MILC value, $f_+(0) = 0.335(36)$, and with the new one obtained by HPQCD, $f_+(0) = 0.332(12)$. All these values are not far from  $f_+(0)=0.304(42)$, often used in the literature and obtained by using the so-called light cone sum rules (LCSR)~\cite{Ball:2004ye}. 

Regarding the total branching fraction, with our new average of the lattice QCD form factor, we obtain: 
\begin{align}
\label{eq:intB-SM-flag}
\mathcal{B}(B\to K & \nu\bar{\nu})^\mathrm{SM}/|\lambda_t|^2 =  
\left\{\begin{matrix}
(1.33 \pm 0.04)_{K_S}\times 10^{-3}\,, &  \\[0.5em]
(2.87 \pm 0.10)_{K^+}\times 10^{-3}\,, & 
\end{matrix}\right.
\end{align}

\noindent where we factored out the CKM dependence and distinguished the charged from the neutral kaon case.~\footnote{Note that the charged mode is also affected by the non-negligible tree-level contribution proportional to $G_F^4$, as discussed in Sec.~\ref{ssec:numerics} of the present letter.} The comparison of this prediction with the ones available in the literature is provided in Table~\ref{tab:form factors}, each corresponding to a different choice of the form factor $f_+(q^2)$: in Ref.~\cite{Buras:2014fpa} the lattice results of Ref.~\cite{Bouchard:2013eph} are combined with the LCSR value; in Ref.~\cite{Blake:2016olu} only the FNAL/MILC lattice results~\cite{Bailey:2015dka} have been used; in Ref.~\cite{Parrott:2022dnu} only the new HPQCD results are considered~\cite{Parrott:2022rgu}; in obtaining our result for the branching fraction we use the average of the 
form factor result obtained by FNAL/MILC~\cite{Bailey:2015dka} and the new HPQCD result~\cite{Parrott:2022rgu}. 
In Fig.~\ref{fig:dBR} we show the impact of the new form factor on the differential branching fraction. In the same plot we emphasize the consequence of using the new form factor average on the error of the branching fraction in the low $q^2$ region.  
One should, however, keep in mind that the form factor at low $q^2$'s is obtained through an extrapolation from large $q^2$'s where the actual lattice QCD data are available. This is done by using a pole-like shape of the form factor, multiplied by a suitable polynomial, as proposed in the model of Ref.~\cite{Bourrely:2008za}. It is therefore of major importance to devise strategies aiming at reducing the impact of hadronic uncertainties in the observables that can be accessed experimentally, as we explore in Sec.~\ref{sec:improved}.

\begin{table}[!t]
\renewcommand{\arraystretch}{1.8}
\centering
\begin{tabular}{|c|c||c|}
\hline 
Ref.  &  Form factors & $\mathcal{B}(B^+\to K^+ \nu\bar{\nu})/|\lambda_t|^{2}$ \\ \hline\hline
Buras et al.~\cite{Buras:2014fpa}  & \cite{Ball:2004ye,Bouchard:2013eph}   &   $(2.5 \pm 0.3)\times 10^{-3}$ \\ 
Blake et al.~\cite{Blake:2016olu}  &  \cite{Bailey:2015dka}  &   $(2.8\pm 0.3)\times 10^{-3}$ \\
Parrott et al.~\cite{Parrott:2022dnu}  &  \cite{Parrott:2022rgu}  &   $(2.81\pm 0.15)\times 10^{-3}$ \\
This work & \cite{Parrott:2022rgu,Bailey:2015dka,Aoki:2021kgd}   &  $(2.87 \pm 0.10)\times 10^{-3}$\\
 \hline
\end{tabular}
\caption{ \sl \small SM predictions for $\mathcal{B}(B^+\to K^+\nu\bar{\nu})/|\lambda_t|^{2}$ and the corresponding form factors used in the computation. Note that the non-negligible contribution proportional to $G_F^4$ has not been included in the above results, cf. Sec.~\ref{ssec:numerics}. }
\label{tab:form factors} 
\end{table}

Before continuing, we need to stress that the difference between $\mathcal{B}(B^0\to K_S \nu\bar{\nu})$ and $\mathcal{B}(B^+\to K^+ \nu\bar{\nu})$ is related to the symmetry relation among the matrix elements:
\begin{align}
\langle K_S| \bar b \gamma_\mu s |B^0\rangle &= - \langle K_S| \bar s \gamma_\mu b |\bar B^0\rangle \nonumber\\ 
&= \frac{1}{\sqrt{2}} \langle K^+ | \bar s \gamma_\mu b |\bar B^+\rangle \,,
\end{align}
where we accounted for the Clebsch-Gordan coefficient, so that in the end we have 
\begin{align}
\mathcal{B}(B^0\to K_S\,\nu\bar{\nu})&= \mathcal{B}(\overline B^0\to K_S\,\nu\bar{\nu}) \nonumber\\ 
& = \frac{1}{2}\frac{\tau_{B^0}}{\tau_{B^+}} \mathcal{B}( B^+\to K^+\nu\bar{\nu}) \,,
\end{align}
where, for illustration purposes, we neglect the tiny phase space difference. In our numerics, however, we use the correct masses of the charged and of the neutral kaons. 
Note in particular that it is important to properly account for the $B$-meson lifetimes, because  $\tau_{B^+}/\tau_{B^0} = 1.076(4)$~\cite{ParticleDataGroup:2022pth}.

\subsection{$B\to K^\ast$ form factors}\label{ssec:form-factors-BKst}

The hadronic matrix element entering the $B\to K^\ast \nu\bar{\nu}$ decay can be parameterized as follows~\footnote{The convention used in Eq.~\eqref{eq:BKst-ff} is $\varepsilon_{0123}=+1$.}
\begin{align}
\label{eq:BKst-ff}
\langle \bar{K}^\ast(k) | \bar{s}\gamma_\mu(1&-\gamma_5) b | \bar{B}(p) \rangle = \varepsilon_{\mu\nu\rho\sigma} \varepsilon^{\ast\nu}p^\rho k^\sigma \dfrac{2 V(q^2)}{m_B+m_{K^\ast}}\nonumber\\[0.4em]
&-i\varepsilon_\mu^\ast(m_B+m_{K^\ast})A_1(q^2) \nonumber\\
&+ i(p+k)_\mu (\varepsilon^\ast\cdot q)\dfrac{A_2(q^2)}{m_B+m_{K^\ast}}\\
&+i q_\mu (\varepsilon^\ast\cdot q) \dfrac{2 m_{K^\ast}}{q^2} \left[A_3(q^2)-A_0(q^2)\right]\,, \nonumber
\end{align}
where $\varepsilon_\mu$ is the polarization vector of $K^\ast$, while $V(q^2)$ and $A_{0,1,2,3}(q^2)$ are the form factors. In the above definition we use $A_3(q^2)$, while in Eq.~\eqref{eq:FF} we used $A_{12}(q^2)$. They are both related to $A_1(q^2)$ and $A_2(q^2)$ as: 
\begin{align}
A_3(q^2) =& \dfrac{m_B+m_{K^\ast}}{2 m_{K^\ast}}A_1(q^2)
- \dfrac{m_B-m_{K^\ast}}{2 m_{K^\ast}}A_2(q^2), \nonumber \\[0.4em]
A_{12}(q^2) =& \dfrac{(m_B+m_{K^\ast}) (m_B^2 -m_{K^\ast}^2 -q^2)}{16 m_B m_{K^\ast}^2} A_1(q^2) \nonumber\\
&- \dfrac{\lambda_{K^\ast}}{16 m_B m_{K^\ast}^2 (m_B+m_{K^\ast}) } A_2(q^2), 
\end{align}
so that at $q^2=0$ they satisfy: $8 m_B m_{K^\ast} A_{12}(0) = (m_B^2 - m_{K^\ast}^2) A_0(0)$, and $A_3(0)=A_0(0)$. Since the pseudoscalar form factor $A_0(q^2)$ does not contribute to the decay rate in the massless neutrino limit, three independent form factors are needed to compute $B\to K^\ast\nu\bar{\nu}$, namely $V$, $A_{1}$ and $A_{2}$.

The situation for the $B\to K^\ast$ transition is far more intricate than for $B\to K$ because there are more form factors. Furthermore, the results of only one lattice QCD study at nonzero recoil have been reported so far, with a specific lattice setup~\cite{Horgan:2013hoa}. 
In this paper, we take the results of Ref.~\cite{Bharucha:2015bzk} in which the lattice QCD values from Ref.~\cite{Horgan:2013hoa} were combined with those obtained by using the LCSR. By adopting the form factor parameterizations and inputs from Ref.~\cite{Bharucha:2015bzk}, we obtain 
\begin{align}
\label{eq:intBst-SM-flag}
\mathcal{B}(B\to K^{\ast } \nu\bar{\nu})^\mathrm{SM}/|\lambda_t|^2 = \left\{\begin{matrix}
(5.9 \pm 0.8)_{K^{\ast 0}}\times 10^{-3}\,, &  \\[0.5em]
(6.4 \pm 0.9)_{K^{\ast +}}\times 10^{-3}\,, & 
\end{matrix}\right.
\end{align}

\noindent in good agreement with Ref.~\cite{Blake:2016olu}.~\footnote{Notice that slightly smaller values are obtained in Ref.~\cite{Buras:2022qip} by using the $B\to K^\ast$ form factors provided in Ref.~\cite{Gubernari:2018wyi}.}
We should point out, however, that this result is obviously less robust than the one for $\mathcal{B}(B \to K  \nu\bar{\nu})$. This is strengthening even more our motivation to look for the options that allow one to reduce sensitivity to the form factor uncertainties, see Sec.~\ref{sec:improved}.

\subsection{CKM couplings}\label{ssec:ckm}

\begin{figure}[t!]
\centering
\includegraphics[width=0.98\linewidth]{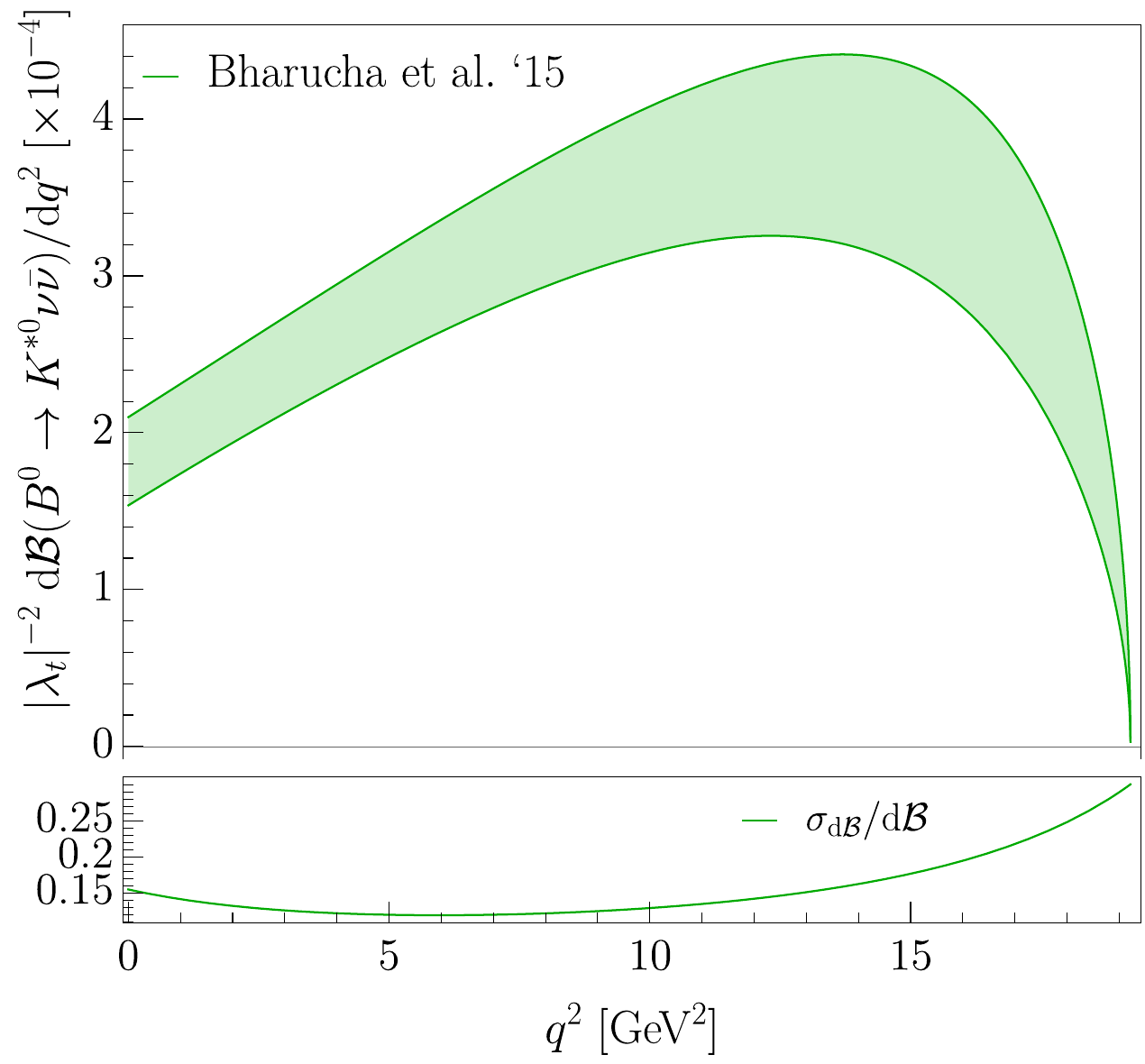}
\caption{\small \sl Predictions for $|\lambda_t|^{-2}\,\mathrm{d}\mathcal{B}(B^0\to K^{\ast 0} \nu\bar{\nu})/\mathrm{d}q^2$ obtained by using the form factors from  Ref.~\cite{Bharucha:2015bzk}. The lower panel shows the relative uncertainty on this quantity as a function of $q^2$. }
\label{fig:dBRst} 
\end{figure}


The uncertainty on the CKM factor $\lambda_t = V_{tb} V_{ts}^\ast$ introduces the largest parametric uncertainty in $\mathcal{B}(B\to K^{(\ast)} \nu\bar{\nu})$. The usual procedure,  often adopted in the literature, is to determine $|V_{cb}|$ from the tree-level processes and then, by virtue of the CKM unitarity, evaluate  $|\lambda_t|$, cf.~e.g.~Ref.~\cite{Buras:2014fpa}. In that way the loop-induced processes are used to probe the effects of physics BSM. Unfortunately, that procedure is intrinsically ambiguous too, because the CKM coupling $|V_{cb}^\mathrm{incl}|$ extracted from the inclusive semileptonic decay, does not coincide with $|V_{cb}^\mathrm{excl}|$ obtained from the exclusive modes. The latest HFLAV average values of the inclusive $|V_{cb}|$ read: $|V_{cb}^\mathrm{incl}|_\mathrm{kin}=(42.2\pm 0.8)\times 10^{-3}$ or $|V_{cb}^\mathrm{incl}|_\mathrm{1S}=(42.0\pm 0.5)\times 10^{-3}$~\cite{HFLAV:2022pwe}. Those values are larger than $|V_{cb}^{B\to D}|=(40.0\pm 1.0)\times 10^{-3}$~\cite{Aoki:2021kgd} obtained after combining the experimental results on the exclusive $B\to D l\nu$ decays~\cite{HFLAV:2022pwe} (with $l=e,\mu$) with the LQCD form factors from Refs.~\cite{MILC:2015uhg,Na:2015kha}. This discrepancy remains true if one compares the inclusive values with the one derived from $B\to D^\ast l\nu$, namely $|V_{cb}^{B\to D^\ast}|=(38.5\pm 0.7)\times 10^{-3}$~\cite{HFLAV:2022pwe}.~\footnote{The quoted $|V_{cb}^{B\to D^\ast}|$ value agrees with $|V_{cb}^{B\to D^\ast}|=(38.9\pm 0.9)\times 10^{-3}$ reported by FLAG~\cite{Aoki:2021kgd}.} In other words, there is a discrepancy in $|\lambda_t|$ depending on the particular input considered,
\begin{equation}
|\lambda_t| \times 10^3=
\left\{\begin{matrix}
41.4 \pm 0.8\,, &\quad\;\, {\small(B\to X_c l \bar{\nu})} \\[0.5em]
39.3 \pm 1.0\,, &\quad {\small(B\to D l \bar{\nu})} \\[0.5em]
37.8 \pm 0.7\,, &\quad\;\, {\small(B\to D^\ast l \bar{\nu})}
\end{matrix}\right.
\end{equation}

\noindent where the inclusive value is about $1\sigma$ and $2\sigma$ larger the the ones derived from $B\to D l \bar{\nu}$ and $B\to D^\ast l \bar{\nu}$ decays, respectively.

For future reference, we will use $|\lambda_t|_\mathrm{excl}=(39.3 \pm 1.0)\times 10^{-3}$, and $|\lambda_t|_\mathrm{incl}$ as noted above. Notice that by using $|\lambda_t|_\mathrm{excl}$ we obtain the branching fractions in Eqs.~(\ref{eq:intB-SM-flag},\ref{eq:intBst-SM-flag}) $1.5\sigma$ smaller than the values we get by using $|\lambda_t|_\mathrm{incl}$, which again highlights the importance of clarifying the issue of the determination of $V_{cb}$.~\footnote{Notice that this discrepancy would be even stronger if instead of $|V_{cb}^{B\to D}|$ we used the HFLAV value of $|V_{cb}^{B\to D^\ast}|$~\cite{HFLAV:2022pwe}. Note also that a larger $|V_{cb}^{B\to D^\ast}|$ has been recently advocated in Ref.~\cite{Martinelli:2021myh}.} 
Another possibility is to rely on the CKM unitarity and extract the $V_{cb}$ from the global fit with data in which $b\to c\ell\nu$ experimental input is removed. This leads to $|V_{cb}^\mathrm{UTfit}| = 42.2(5)\times 10^{-3}$~\cite{UTfit:2022hsi}, or $|V_{cb}^\mathrm{CKMfitter}| = 41.3(3)\times 10^{-3}$~\cite{Charles:2004jd}, mostly constrained by the $K^0-\overline K^0$ and $B_s - \overline B_s$ mixing. Clearly, one could extract $\lambda_t$ from $\Delta m_{B_s}$, as advocated in Ref.~\cite{Buras:2022qip}. In all these alternative strategies, an excellent control over the matrix element of the $B_s - \overline B_s$ mixing is required, usually referred to as $f_{B_s}\sqrt{\hat B_{B_s}}$, which is currently not the case. In the FLAG review, $f_{B_s}\sqrt{\hat B_{B_s}}= 274(8)$~MeV ($N_\mathrm{f}=2+1$), and $f_{B_s}\sqrt{\hat B_{B_s}}= 256(6)$~MeV ($N_\mathrm{f}=2+1+1$), which again renders the value of $\lambda_t$ ambiguous. One way to avoid the ambiguity on the CKM couplings will be discussed in Sec.~\ref{sec:improved}.

\subsection{Numerical Predictions}\label{ssec:numerics}
Before we give our final numerical results, we need to account for one significant correction. It was first noted in Ref.~\cite{Kamenik:2009kc} that there is an important tree level contribution to the charged $B^\pm\to K^\pm\nu\bar{\nu}$ mode arising from the weak annihilation mediated by the on-shell $\tau$-lepton. We will follow Ref.~\cite{Kamenik:2009kc} and call this new contribution as ``Tree", while keeping in mind that its amplitude is $\propto G_F^2$. To be more specific, we write:
\begin{align}
\label{eq:treeDD}
   { d\mathcal{B}(B^+\to K^+\nu\bar{\nu})_\mathrm{Tree}\over dq^2 dp_\tau^2}
&=  {\tau_{B^+} G_F^4 |V_{us}V_{ub} |^2 f_K^2 f_B^2 \over 64 \pi^3 m_B^3}\, p_\tau^4 \nonumber\\
&\hspace{-19mm} \times \frac{m_B^2 (p_\tau^2 -m_K^2) -  p_\tau^2 (p_\tau^2 +q^2 -m_K^2)}{ 
(m_\tau^2 -p_\tau^2)^2 + m_\tau^2 \Gamma_\tau^2}\,,
\end{align}
where $p_\tau^2$ and $\Gamma_\tau$ are the invariant mass and the full width of the intermediate $\tau$-lepton. After integrating over the phase space we have:
\begin{align}
    \mathcal{B}(B^+\to K^+\nu\bar{\nu})_\mathrm{Tree} 
&=  {\tau_{B^+} G_F^4 |V_{us}V_{ub}|^2 f_K^2 f_B^2 \over 128 \pi^2 m_B^3}\,  \nonumber\\
&\hspace{-10mm} \times \frac{m_\tau}{\Gamma_\tau} (m_\tau^2 -m_K^2)^2 (m_B^2 - m_\tau^2)^2\,,
\end{align}
where we have used a narrow-width approximation. Similarly, for the $B^+\to K^{\ast +}\nu\bar{\nu}$ mode we have,
\begin{align}
\label{eq:treeDDsst}
 { d\mathcal{B}(B^+\to K^{\ast +}\nu\bar{\nu})_\mathrm{Tree}\over dq^2 dp_\tau^2}
&=  {\tau_{B^+} G_F^4 |V_{us}V_{ub} |^2 f_{K^\ast}^2 f_B^2 \over 64 \pi^3 m_B^3}\, p_\tau^4 \nonumber \\ 
&\hspace{-26mm} \times \frac{(m_B^2 - p_\tau^2) (p_\tau^2 -m_{K^\ast}^2) -  q^2 (p_\tau^2 - 2m_{K^\ast}^2)}{ 
(m_\tau^2 -p_\tau^2)^2 + m_\tau^2 \Gamma_\tau^2}\,,
\end{align}
which leads to,
\begin{align}
& \mathcal{B}(B^+\to K^{\ast +}\nu\bar{\nu})_\mathrm{Tree} 
=  {\tau_{B^+} G_F^4 |V_{us}V_{ub}|^2 f_{K^\ast}^2 f_B^2 \over 128 \pi^2 m_B^3}  \frac{m_\tau}{\Gamma_\tau}  \nonumber\\
&\quad\,\,\,\,\times (m_\tau^2 -m_{K^\ast}^2)^2 (m_B^2 - m_\tau^2)^2 \left( 1 + \frac{2 m_{K^\ast}^2}{m_\tau^2}\right)\,.
\end{align}
To estimate the size of this contribution we take  $f_K=155.7(3)$~MeV and $f_B=190.0(1.3)$~MeV from the FLAG review~\cite{Aoki:2021kgd}, in addition to  $f_{K^\ast}=205(6)$~MeV, which we extracted from the measured $\mathcal{B}(\tau\to K^\ast \nu)= 1.20(7)\%$~\cite{ParticleDataGroup:2022pth}, and by using $\vert V_{us}\vert = 0.2259(5)$~\cite{Becirevic:2020rzi}, $\vert V_{ub}\vert = 3.74(17) \times 10^{-3}$~\cite{Aoki:2021kgd}. We finally obtain 
\begin{align}
\begin{split}
\mathcal{B}(B^+\to K^+\nu\bar{\nu})_\mathrm{Tree} &= (6.28\pm 0.06)\times 10^{-7} 
\,, \\[0.35em]
\mathcal{B}(B^+\to K^{\ast +}\nu\bar{\nu})_\mathrm{Tree} &= (1.07\pm 0.10)\times 10^{-6} \,.
\end{split}
\end{align}
Despite being $\propto G_F^4$ this contribution is indeed numerically significant. We find that it amounts to more than $10\%$ with respect to the dominant effect $\propto G_F^2$, and therefore if we aim at the $10\%$ experimental precision or better at Belle-II, the ``Tree" contribution must be included~\cite{Belle-II:2018jsg}. We do so to obtain our final estimates for the total branching fractions collected in Table.~\ref{tab:final-predictions}, which are just below the current experimental limits~\cite{Belle:2017oht}. Note, once again, that in our final results we used 
$|\lambda_t|=(3.93\pm 0.10)\times 10^{-2}$. 
For the readers' convenience, in Tables~\ref{tab:binned-predictions-K} and \ref{tab:binned-predictions-Kst} we also provide the binned values for the $B\to K^{(\ast)} \nu \bar{\nu}$ branching fractions.
\begin{table}[!t]
\renewcommand{\arraystretch}{1.9}
\centering
\begin{tabular}{|c|c|}
\hline 
Decay & Branching ratio \\ \hline\hline
$B^+\to K^+ \nu \bar{\nu}$ & $(5.06 \pm 0.14 \pm 0.28
)\times 10^{-6}$\\
$B^0\to K_S \nu \bar{\nu}$ & $(2.05 \pm 0.07 \pm 0.12)\times 10^{-6}$ \\ 
$B^+\to K^{\ast +} \nu \bar{\nu}$ & $(10.86 \pm 1.30 \pm 0.59)\times 10^{-6}$ \\
$B^0\to K^{\ast 0} \nu \bar{\nu}$ & $(9.05 \pm 1.25 \pm 0.55)\times 10^{-6}$  \\ \hline
\end{tabular}
\caption{ \sl \small Our final predictions for the branching fractions of the most relevant $b\to s\nu\bar \nu$ decay modes. The first uncertainty comes from the hadronic form factors and the second one is dominated by the uncertainty of $|\lambda_t|$. }
\label{tab:final-predictions} 
\end{table}
\begin{table*}[]
\renewcommand{\arraystretch}{1.9}
\centering
\begin{tabular}{|c||c|c||c|c|}
\hline 
$q^2$-bin $[\mathrm{GeV}^2]$ &  $\mathcal{B}(B^+ \to K^+ \nu\bar{\nu})\times 10^{6}$ & $\sigma_{\mathcal{B}_{K^+}}/\mathcal{B}_{K^+}$&  $\mathcal{B}(B^0 \to K_S \nu\bar{\nu})\times 10^{6}$ & $\sigma_{\mathcal{B}_{K_S}}/\mathcal{B}_{K_S}$ \\ \hline\hline
$[0,4]$ & $(1.206 \pm 0.055 \pm 0.066)$ & 0.07 & 
 $(0.490 \pm 0.026 \pm 0.030)$ & 0.08 \\ 
$[4,8]$ & $(1.161 \pm 0.039 \pm 0.064)$ & 0.06 &
 $(0.477 \pm 0.018 \pm 0.029)$ & 0.07  \\ 
$[8,12]$ & $(1.064\pm 0.027 \pm 0.059)$ & 0.06 &
$(0.439\pm 0.013 \pm 0.027)$ & 0.07  \\ 
$[12,16]$ & $(0.889 \pm 0.020 \pm 0.049)$ & 0.06 & 
 $(0.365 \pm 0.009 \pm 0.022)$ & 0.07  \\ 
$[16,q^2_\mathrm{max}]$ & $(0.744 \pm 0.017 \pm 0.039 )$ & 0.06 &
$(0.282 \pm 0.008 \pm 0.017)$ & 0.07 \\[0.3em] \cdashline{1-5}
$[0,q^2_\mathrm{max}]$ & $(5.06 \pm 0.14 \pm 0.28)$ & 0.06&
$(2.05 \pm 0.07 \pm 0.12)$ & 0.07  \\ \hline

\end{tabular}
\caption{ \sl \small SM predictions for the partially integrated $B\to K \nu\bar{\nu}$ branching fraction, in a given $q^2$-bin, obtained by using our fit to the lattice form factors from HQPCD~\cite{Parrott:2022rgu} and FNAL/MILC~\cite{Bailey:2015dka}, and the CKM input~$|\lambda_t|=3.93(10)\times 10^{-2}$. The first uncertainty comes from the hadronic form factors and the second one is dominated by the uncertainty on $|\lambda_t|^2$. The total relative uncertainty of each observable is shown in the last column. }
\label{tab:binned-predictions-K} 
\end{table*}
\begin{table*}[]
\renewcommand{\arraystretch}{1.9}
\centering
\begin{tabular}{|c||c|c||c|c|}
\hline 
$q^2$-bin $[\mathrm{GeV}^2]$ & $\mathcal{B}(B^+ \to K^{\ast+} \nu\bar{\nu})\times 10^{6}$ & $\sigma_{\mathcal{B}_{K^{\ast +}}}/\mathcal{B}_{K^{\ast+}}$&$\mathcal{B}(B^0 \to K^{\ast0} \nu\bar{\nu})\times 10^{6}$ & $\sigma_{\mathcal{B}_{K^{\ast 0}}}/\mathcal{B}_{K^{\ast 0}}$\\ \hline\hline
$[0,4]$ &$(1.77 \pm 0.20 \pm 0.09)$ & 0.12 &
$(1.38 \pm 0.18 \pm 0.08)$ & 0.15 \\ 
$[4,8]$ & $(2.25 \pm 0.23 \pm 0.12)$  & 0.10
&$(1.85 \pm 0.22 \pm 0.11)$  & 0.13 \\ 
$[8,12]$ 
&$(2.63 \pm 0.30 \pm 0.15)$  &0.13
&$(2.23 \pm 0.28 \pm 0.14)$  &0.14
 \\ 
$[12,16]$ 
& $(2.71 \pm 0.39 \pm 0.15 )$ &0.15
& $(2.32 \pm 0.37 \pm 0.14)$ &0.17
\\ 
$[16,q^2_\mathrm{max}]$ 
&  $(1.50 \pm 0.30 \pm 0.09)$ &0.21
&  $(1.27 \pm 0.29 \pm 0.08)$ &0.23
\\[0.3em] \cdashline{1-5}
$[0,q^2_\mathrm{max}]$ 
&$(10.86 \pm 1.30 \pm 0.59)$  &0.12
&$(9.05 \pm 1.25 \pm 0.55)$ &0.15\\ \hline

\end{tabular}
\caption{ \sl \small SM predictions similar to those presented in Tab.~\ref{tab:binned-predictions-K} but for the case of vector meson in the final state, $B\to K^\ast \nu\bar{\nu}$. }
\label{tab:binned-predictions-Kst} 
\end{table*}

Before closing this section, we should stress that depending on the precision the so-called ``Tree" contribution can be important when discussing the ratio $r_\mathrm{lh}$~\eqref{eq:rlh}. While the case of $B \to K_S \nu \bar{\nu}$ remains unchanged, we note that for the charged mode:
\begin{align}
   & \frac{\mathcal{B}(B^+\to K^+ \nu \bar{\nu})_\mathrm{Tree}}{\mathcal{B}(B^+\to K^+ \nu \bar{\nu})} \approx \biggl. 11\,\% \biggr|_{\mathrm{low}-q^2},\, \biggl. 13\%\biggr|_{\mathrm{high}-q^2},
\end{align}
 where the denominator represents the sum of values obtained by using Eq.~\eqref{eq:dBR} and Eq.~\ref{eq:treeDD}.

\section{Improved strategies}\label{sec:improved}
We now explore the strategies allowing us to reduce the impact of hadronic uncertainties, while trying to keep the sensitivity to the BSM physics pronounced. In the following discussion, we focus on 
$B\to K_S\,\nu\bar{\nu}$ which is not impacted by the above mentioned ``Tree" contribution.

\subsection{$B\to K \nu\bar{\nu}$ at high-$q^2$}\label{ssec:binned}

For $B\to K \nu\bar{\nu}$ decays, the simplest strategy to reduce the theoretical uncertainty is to focus on the high-$q^2$ region, which is where $f_+(q^2)$ is precisely determined in LQCD. This restriction to high $q^2$'s does hamper the sensitivity of this quantity to New Physics which would only rescale the entire $q^2$-spectrum for left-handed neutrinos (see Sec.~\ref{sec:bsm}).~\footnote{See Ref.~\cite{Felkl:2021uxi} for a discussion that includes right-handed neutrinos too.} The main disadvantage of this approach is that less statistics would be available in the experimental measurement, which to our view is outweighed by the advantage of having better theoretical control over the SM prediction. 

By considering the two intervals, $q^2 \geq 12~\mathrm{GeV}^2$ (bin I) and $q^2 \geq 16~\mathrm{GeV}^2$ (bin II), we find that they comprise about $30\,\%$ and $15\,\%$ of the full event sample, respectively. The corresponding SM predictions in these bins are: 
\begin{small}
\begin{align}
\mathcal{B}(B\to K_S \nu\bar{\nu})^\mathrm{SM}_{\text{bin~I}} &=
(0.647 \pm 0.017 \pm 0.039)\times 10^{-6}\,,  \\[0.45em]
\mathcal{B}(B\to K_S \nu\bar{\nu})^\mathrm{SM}_{\text{bin~II}} &=
(0.282 \pm 0.008 \pm 0.017)\times 10^{-6}\,,
\end{align}
\end{small}

\noindent where the form factor uncertainties become subdominant in comparison with the one arising from the CKM matrix elements. Note also that the relative uncertainties in these intervals are about a factor of $2$ smaller than the one from the total branching fraction, see Table~\ref{tab:final-predictions}. 

This strategy provides a clear way to avoid the uncontrolled form factor uncertainties in the $B\to K$ transition, provided the binned information will be made available by Belle-II. However, there are several limitations to this approach. Firstly, the uncertainty associated with the CKM factor remains important, in particular the one arising from discrepancies between different determinations of  $|V_{cb}|$, cf. Sec.~\ref{ssec:ckm}. Moreover, as of now, this idea cannot be fully exploited for $B\to K^\ast$ decays since the LQCD results at nonzero recoil have only been obtained by one collaboration and without a full control over systematic uncertainties~\cite{Horgan:2013hoa}. These limitations call for the alternative approaches to reduce the theoretical errors, as discussed in the following.

\subsection{$(\nu/\ell)$ ratios}\label{ssec:ratios}

In this Section we explore an alternative way to reduce the theoretical uncertainty on both $B\to K\nu\bar{\nu}$ and $B\to K^\ast\nu\bar{\nu}$ decays by exploiting the similarity with the corresponding decays into charged leptons. More precisely, we study the following ratio,
\begin{align}
\label{eq:nu-mu-ratio}
\mathcal{R}_{K^{(\ast)}}^{(\nu/l)} [q_0^2,q_1^2]\equiv \dfrac{\mathcal{B}(B\to K^{(\ast)} \nu\bar{\nu})}{\mathcal{B}(B\to K^{(\ast)} ll)}\Bigg{|}_{ [q_0^2,q_1^2]}\,,
\end{align}

\noindent where $l \in \lbrace e,\mu \rbrace$, and the branching fractions are integrated over the same $q^2$ interval, $[q_0^2,q_1^2]$, both in the numerator and in the denominator. Since the considered lepton masses are negligible with respect to the other mass scales in the process, we can expect a cancellation not only of the CKM factors, but also of the form factors in Eq.~\eqref{eq:nu-mu-ratio}, provided the $q^2$-bin is chosen judiciously. Of course, the region around the $c\bar{c}$ resonances must be avoided, as the resonances would completely spoil the benefits of the ratio $\mathcal{R}_{K^{(\ast)}}^{(\nu/l)} [q_0^2,q_1^2]$. Moreover, one should consider a region where $C_9^\mathrm{eff}(q^2)$ is under reasonable theoretical control. The optimal choice turns out to be the interval $q^2\in [1.1,6]~\mathrm{GeV}^2$, the one that is already considered in the experimental tests of lepton flavor universality (LFU)~\cite{LHCb:2022qnv}.

We briefly remind the reader that the processes based on $b\to s\ell\ell$ can be described by the following effective Lagrangian~\cite{Altmannshofer:2008dz}:
\begin{align}
\label{eq:eft-bsll}
   \hspace{-0.4em} \mathcal{L}_\mathrm{eff}^{b\to s\ell\ell} = \dfrac{4 G_F}{\sqrt{2}} \lambda_t \sum_i \Big{(} C_i^\ell \,\mathcal{O}_i^\ell +  C_{i^\prime}^{\ell}\,\mathcal{O}_{i^\prime}^{
    \ell} \Big{)}+\mathrm{h.c.}\,,
\end{align}
where the effective coefficients $C_i^{\ell\ell} \equiv C_i^{\ell\ell}(\mu)$ and operators $\mathcal{O}_i^{\ell\ell} \equiv \mathcal{O}_i^{\ell\ell}(\mu)$ are defined at the scale $\mu=m_b$. The relevant operators to our study are
\begin{align}
\mathcal{O}_9^{\ell\ell} &= \dfrac{e^2}{(4\pi)^2} \big{(}\bar{s}\gamma_\mu P_L b \big{)}\big{(}\bar{\ell}\gamma_\mu \ell \big{)}\,,\\
\mathcal{O}_{10}^{\ell\ell} &= \dfrac{e^2}{(4\pi)^2} \big{(}\bar{s}\gamma_\mu P_L b \big{)}\big{(}\bar{\ell}\gamma_\mu\gamma_5 \ell \big{)}\,,
\end{align}
in addition to the chirality flipped ones, $\mathcal{O}_{i^\prime}^{\ell\ell}$, obtained from $\mathcal{O}_i^{\ell\ell}$ by replacing $P_L \leftrightarrow P_R$, and the dipole operators $\mathcal{O}_{7,8}$. The contributions from the four-quark operators $\mathcal{O}_{1-6}$ are included in the redefinition of the coefficients $C_{7,9}$~\cite{Altmannshofer:2008dz,Buras:1993xp,Bobeth:1999mk}. 

\begin{figure}[t!]
\centering
\includegraphics[width=.98\linewidth]{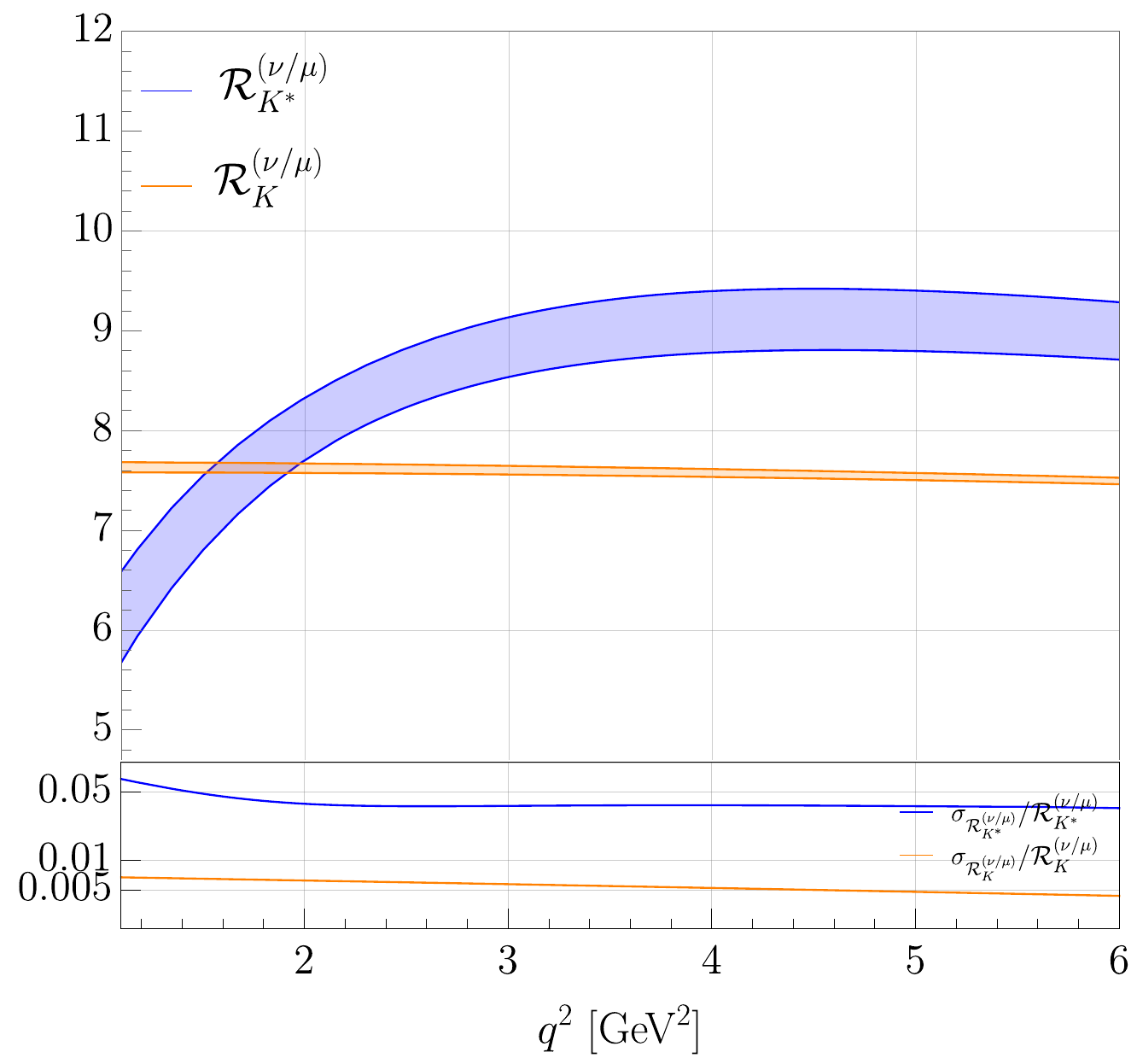}
\caption{\small \sl The ratios $\mathcal{R}_{K}^{(\nu/\mu)}$ and $\mathcal{R}_{K^{\ast}}^{(\nu/\mu)}$ are plotted in orange and blue respectively, as  functions of $q^2$ in the region well below the first $c\bar{c}$ resonance. In the lower panel we display the relative error in order to better appreciate the precision with respect to those shown in Figs.~\ref{fig:dBR} and~\ref{fig:dBRst}. 
}
\label{fig:ratios-nu-l-SM} 
\end{figure}

By relying on the theoretical inputs described above, we obtain the following predictions for the $B\to K^{(\ast)}ll$ partial branching fractions (with $l=e,\mu$),
\begin{align}
\mathcal{B}(B^0 \to K_S ll)^\mathrm{SM}_{[1.1,6]}/|\lambda_t|^2 &=0.507(24)\times 10^{-4} \,,\\[0.4em]
\mathcal{B}(B^0 \to K^{\ast 0}ll)^\mathrm{SM}_{[1.1,6]}/|\lambda_t|^2 &=1.46(21)\times 10^{-4} \,.
\end{align}
The difference between the rates with muons and with electrons in the final state is completely marginal with respect to the current theoretical uncertainties. By combining these results with $\mathcal{B}(B\to K^{(\ast)}\nu\bar \nu)_{[1.1,6]}$, we find: 
\begin{align}
\label{eq:RnulK}
\biggl.\mathcal{R}_{K }^{(\nu/l)}[1.1,6]\biggr|_\mathrm{SM} &=7.58\pm 0.04\, , 
\end{align}
\noindent and 
\begin{align}
\label{eq:RnulKstar}
\biggl. \mathcal{R}_{K^\ast}^{(\nu/l)}[1.1,6]\biggr|_\mathrm{SM} &=  8.6\pm 0.3 \,,
\end{align}
\noindent which are shown in Fig.~\ref{fig:ratios-nu-l-SM}. For the ratios based on $B\to K$ decays, we obtain a relative uncertainty less than $1\%$, which is much smaller the $7.5\%$ error on $\mathcal{B}(B\to K \nu\bar \nu)_{[1.1,6]}$. The cancellation of the $f_+(q^2)$ form factor in $\smash{\mathcal{R}_{K}^{(\nu/l)}}$ is indeed efficient, since the contributions involving the scalar ($f_0$) and tensor ($f_T$) form factors are suppressed by $m_l^2/q^2$ and by $|C_7|/|C_{9}| \approx 0.1$, respectively. The same argument holds true for the pseudoscalar ($A_0$) and the tensor form factors ($T_{1,2,3}$)  entering the $B\to K^\ast$ observables. However, in the latter case, there are more form factors that survive even in the limit in which the charged lepton mass and the photon-dipole are neglected. Therefore the cancellation of the form factors work to a lesser extent and the ratio $\smash{\mathcal{R}_{K^\ast}^{(\nu/l)}}$ is predicted with a relative uncertainty of about $5\%$ in the bin we chose, $q^2\in [1.1,6]~\mathrm{GeV}^2$. This resulting uncertainty is, however, almost a factor $3$ smaller than the one in $\mathcal{B}(B\to K^\ast \nu\bar{\nu})_{[1.1,6]}$, proving that the proposal to study the ratio could be advantegous also in this case.

\subsection{Standard Model Considerations}

The caveat in the above discussion is related to the choice of $C_9^\mathrm{eff}(q^2)$ that one has to make in order to evaluate the full and/or partial branching fractions, $\mathcal{B}(B\to K^{(\ast )} ll)$. That issue attracted quite a lot of attention in the literature and we will not dwell on it here. To obtain the results in Eqs.~(\ref{eq:RnulK},\ref{eq:RnulKstar}) we relied on the quark-hadron duality and used the two-loop results from Ref.~\cite{Asatryan:2001zw}. The inclusion of the non-factorizable nonperturbative $c\bar c$-contribution to $C_9^\mathrm{eff}$ has been discussed in Ref.~\cite{Khodjamirian:2012rm}, and recently improved in Refs.~\cite{Gubernari:2020eft,Gubernari:2022hxn}. A reliable estimate of that contribution remains a challenging task. In practice, for example, the quark-hadron duality, mentioned above, in the interval $q^2\in [1.1,6]~\mathrm{GeV}^2$, leads to $C_9^\mathrm{eff}\simeq 4.4$, also used in the popular {\tt flavio} code~\cite{Straub:2018kue}. Instead, in Ref.~\cite{Ciuchini:2020gvn} the value $C_9^\mathrm{eff}\simeq 4.1$ is preferred. This is actually the main source of disagreement among various collaborations while performing the global analyses of the experimental data for the angular observables in the exclusive $b\to sll$ decay modes~\cite{Ciuchini:2020gvn,Alguero:2019ptt}. 

\begin{figure}[t!]
\centering
\includegraphics[width=1.\linewidth]{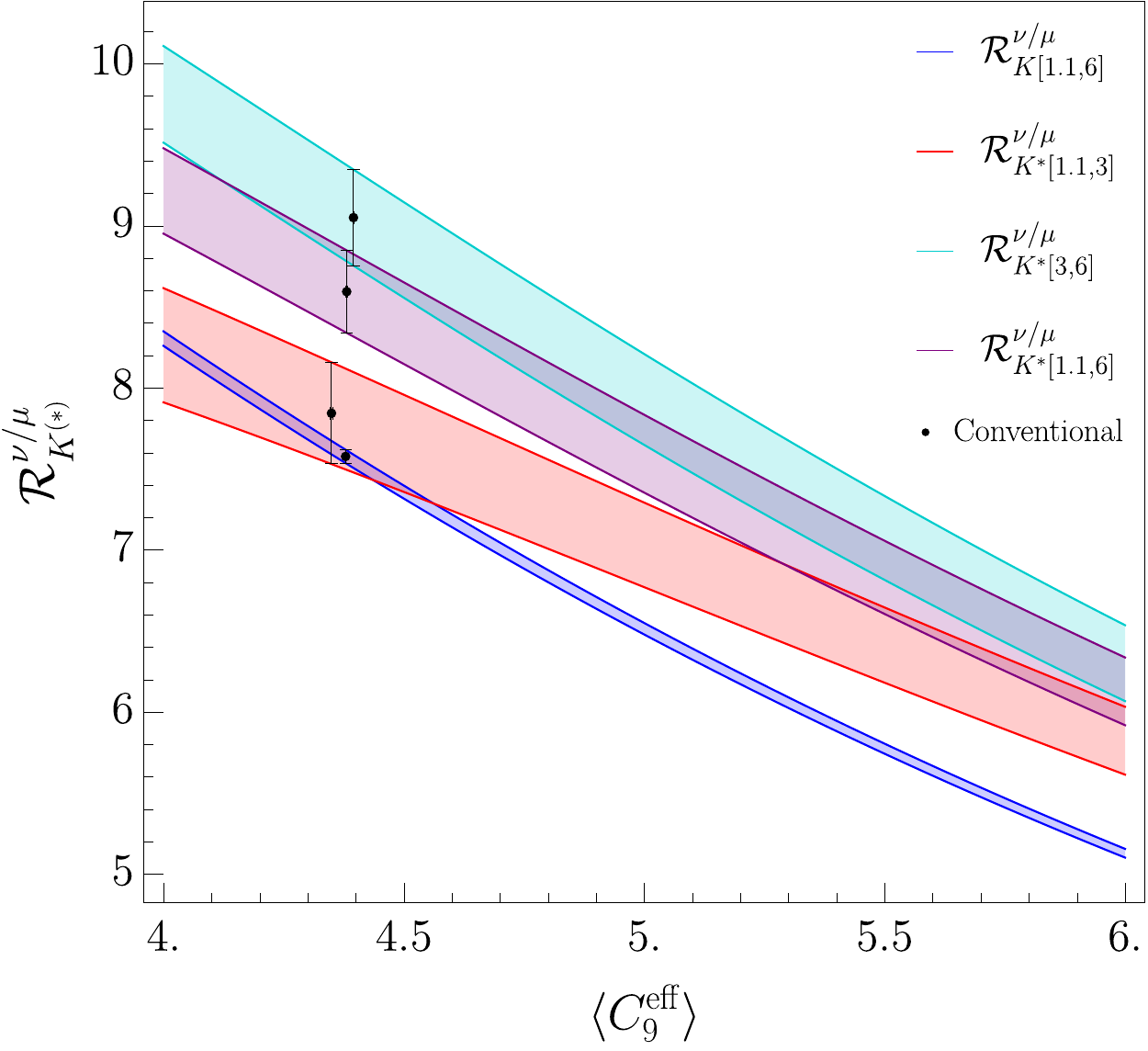}
\caption{\small \sl The ratios $\mathcal{R}_{K}^{(\nu/\mu)}$ and $\mathcal{R}_{K^{\ast}}^{(\nu/\mu)}$ are plotted as a function of the average value of $C_9^\mathrm{eff}$ in a  fixed $q^2$ bin. The black points correspond to the predictions obtained using the $C_9^\mathrm{eff}$ input from Ref.~\cite{Straub:2018kue}. Note that the ratio $\mathcal{R}_{K}^{(\nu/\mu)}$ is practically constant in the low-$q^2$ region and, for this reason, is only plotted in the $[1.1,6]~\mathrm{GeV}^2$ bin, see Fig.~\ref{fig:ratios-nu-l-SM}  .}
\label{fig:ratios-nu-l-SM-bis} 
\end{figure}

We can now turn that problem around and actually use our ratios $\mathcal{R}_{K^{(\ast )}}^{(\nu/e)}[1.1,6]$ in order to extract the $C_9^\mathrm{eff}$ from the data. If we stick to the SM, we obtain 
\begin{align}
\label{eq:RnulKC9}
\biggl.\frac{1}{\mathcal{R}_{K }^{(\nu /l)}[1.1,6]}\biggr|_\mathrm{SM} &\simeq \nonumber \\
&\hspace*{-12mm}\left[7.15 -0.45 \cdot C_9^\mathrm{eff} + 0.42\cdot  \left(C_9^\mathrm{eff}\right)^2 \right] \times 10^{-2}\,,\\[0.4em]
\label{eq:RnulKstarC9}
\biggl.\frac{1}{\mathcal{R}_{K^\ast }^{(\nu /l)}[1.1,6]}\biggr|_\mathrm{SM} &\simeq \nonumber \\
 &\hspace*{-12mm} \left[
 9.98 -1.45 \cdot C_9^\mathrm{eff} +0.42 \cdot  \left(C_9^\mathrm{eff}\right)^2
 \right] \times 10^{-2}\,, 
\end{align}
where we neglect the $q^2$-variation of $C_9^\mathrm{eff}$, which is why in Fig.~\ref{fig:ratios-nu-l-SM-bis} we employed $\langle C_9^\mathrm{eff}\rangle$ to emphasize that it corresponds to the average of $C_9^\mathrm{eff}(q^2)$ over the interval of $q^2$'s in which the ratio $\mathcal{R}_{K^{(\ast )}}^{(\nu /l)}$ has been measured. 
It is interesting to note that if $\mathcal{R}_{K}^{(\nu /l)}[1.1,6]$ is measured with a precision of $10\%$ ($20\%$), the error on the extracted $\langle C_9^\mathrm{eff}\rangle$ would be $\pm 0.4$ ($\pm 0.8$). 
Finally, we should stress that it is important to consider both the pseudoscalar and the vector kaon in the final state, since the non-factorizable contributions are expected to be smaller in the case of $K$ than in the case of outgoing $K^\ast$~\cite{Khodjamirian:2012rm}. 

\section{BSM implications}\label{sec:bsm}

In this Section we discuss the sensitivity of the observables discussed above to the contributions arising from physics BSM. We will provide the most general expressions for the ratios defined in Eq.~\eqref{eq:nu-mu-ratio} and briefly illustrate their variation with respect to the SM in the case of a peculiar New Physics scenario.

\paragraph*{General expressions:} We factorize the SM contribution and, in a given $q^2$-bin, we write:
\begin{align}
\mathcal{R}_{K^{(\ast)}}^{(\nu/l)}  = \mathcal{R}_{K^{(\ast)}}^{(\nu/l)}\Big{|}_\mathrm{SM} \left(1+ \delta \mathcal{R}^{(\nu/l)}_{K^{(\ast)}} \right)\,, 
\end{align}
where $\smash{\delta \mathcal{R}^{(\nu/l)}_{K^{(\ast)}}}$ is the New Physics contribution. Similarly, $\delta \mathcal{B}_{K^{(\ast)}}^{\nu\bar{\nu}}$ and $\delta\mathcal{B}_{K^{(\ast)}}^{ll}$ are the respective shifts of $\mathcal{B}(B\to K^{(\ast)}\nu\bar{\nu})$ and $\mathcal{B}(B\to K^{(\ast)}ll)$ in a given $q^2$ bin, arising from BSM, 
{
\small
\begin{align}
\begin{split}
\mathcal{B}(B\to {K^{(\ast)}}{\nu\nu}) &= \mathcal{B}(B\to {K^{(\ast)}}{\nu\nu})\Big{|}_\mathrm{SM} \, \big{(}1+\delta \mathcal{B}_{K^{(\ast)}}^{\nu\nu}\big{)}\,,\\[0.35em]
\small\mathcal{B}(B\to {K^{(\ast)}}{ll}) &= \mathcal{B}(B\to {K^{(\ast)}}{ll})\Big{|}_\mathrm{SM} \, \big{(}1+\delta \mathcal{B}_{K^{(\ast)}}^{ll}\big{)}\,.
\end{split}
\end{align}
}
\noindent In other words, we can write:

\begin{align}
\label{eq:ratio-magic-numbers}
\delta \mathcal{R}_{K^{(\ast)}}^{(\nu/l)} = \displaystyle\dfrac{\delta \mathcal{B}_{K^{(\ast)}}^{\nu\bar{\nu}}-\delta \mathcal{B}_{K^{(\ast)}}^{ll}}{1+\delta \mathcal{B}_{K^{(\ast)}}^{ll}}\,.
\end{align}
The $\delta \mathcal{B}_{K^{(\ast)}}^{\nu\bar{\nu}}$ can be easily expressed in terms of the Wilson coefficients defined in the Lagrangian~\eqref{eq:eft-bsnunu}, namely,
\begin{align}
\begin{split}
\delta \mathcal{B}_{K^{(\ast)}}^{\nu\bar{\nu}} &=  \sum_{i}\dfrac{2\mathrm{Re}[C_L^\mathrm{SM}\,(\delta C_{L}^{\nu_i\nu_i}+\delta C_{R}^{\nu_i\nu_i})]}{3|C_{L}^\mathrm{SM}|^2}\\
&+\sum_{i,j}\dfrac{|\delta C_{L}^{\nu_i\nu_j}+\delta C_{R}^{\nu_i\nu_j}|^2}{3|C_L^\mathrm{SM}|^2}\\
&- \eta_{K^{(\ast)}}\sum_{i,j} \dfrac{\mathrm{Re}[\delta C_R^{\nu_i\nu_j}(C_{L}^\mathrm{SM}\delta_{ij}+\delta C_{L}^{\nu_i\nu_j})]}{3|C_{L}^\mathrm{SM}|^2}\,,
\end{split}
\end{align}

\begin{table*}[t]
\renewcommand{\arraystretch}{1.8}
\centering
\begin{tabular}{|c|cccc|cccc|}
\hline 
Decays ($l=e,\mu$) & $a_{VV}$ & $a_{VA}$ & $a_{AV}$ & $a_{AA}$ & $b_{VV}$ & $b_{VA}$ & $b_{AV}$ & $b_{AA}$\\\hline \hline
$B\to K ll$ & $0.2430(1)$ & $-0.260(1) $& $0$ & $0$ & $0.0316(2)$ & $0.0317(2)$ & $0$ & $0$ \\ 
$B\to K^\ast ll$ & $0.0012(48)$  & $-0.038(8)$ & $-0.191(10)$ & $0.255(6)$ & $0.0048(10)$ & $0.0047(10)$ & $0.0312(7)$ & $0.0311(7)$ \\ \hline
\end{tabular}
\caption{ \sl \small Numerical coefficients defined in Eq.~\eqref{eq:magic-BKll} for $l=e,\mu$, which are computed in the $[1.1,6]~\mathrm{GeV}^2$ bin by using the $B\to K$ and $B\to K^\ast$ form factors from Ref.~\cite{Aoki:2021kgd} and Ref.~\cite{Bharucha:2015bzk}, respectively. The difference between electron and muon coefficients lies within the quoted uncertainties. }
\label{tab:magic}  
\end{table*}

\noindent where we sum over the neutrino flavors $i,j \in \lbrace 1,2,3 \rbrace$. Note that the last term vanishes for $B\to K\nu\bar{\nu}$ ($\eta_K=0$), but it is nonzero for $B\to K^\ast\nu\bar{\nu}$. Using the form factor described in Sec.~\ref{sec:BKnunu}, we find that $\eta_{K^\ast}=3.47(10)$ in the $[1.1,6]~\mathrm{GeV}^2$ bin. Similarly for $\delta\mathcal{B}_{K^{(\ast)}}^{ll}$ we have,
\begin{align}
\label{eq:magic-BKll}
\delta\mathcal{B}_{K^{(\ast)}}^{ll} &= \sum_i a_i \,\mathrm{Re}\big{(}\delta C_i^{ll}\big{)} + \sum_i b_i \,|\delta C_i^{ll}|^2
\end{align}
\noindent where $i \in \lbrace {VV},{AV},{VA},{AA} \rbrace$, with
\begin{align}
\begin{split}
\delta C_{VV}^{ll} &= \delta C_{9^\prime}^{ll} + \delta C_{9}^{ll}\,,  \qquad \delta C_{VA}^{ll} = \delta C_{10^\prime}^{ll} + \delta C_{{10}}^{ll}\,,\\[0.4em]
\delta C_{AV}^{ll} &= \delta C_{9^\prime}^{ll} - \delta C_{9}^{ll}\,, \qquad \delta C_{AA}^{ll} = \delta C_{10^\prime}^{ll} - \delta C_{{10}}^{ll}\,.
\end{split}
\end{align}
We computed the numerical coefficients $a_i$ and $b_i$, and the results are collected in Table~\ref{tab:magic}.

\paragraph*{SMEFT:} Since we are interested in New Physics scenarios defined well above the electroweak scale, the low-energy effective theory must be replaced by the SM effective field theory (SMEFT), which is invariant under the full $SU(3)_c\times  SU(2)_L \times U(1)_Y$ gauge symmetry~\cite{Buchmuller:1985jz}. The main interest of using this approach is that contributions to the $b\to s\nu\bar{\nu}$ are partially correlated to $b\to s\mu\mu$ via $SU(2)_L$ gauge invariance~\cite{Buras:2014fpa,Bause:2021cna,deGiorgi:2022vup,Rajeev:2021ntt}.

The $d=6$ semileptonic operators relevant to our study are:
\begin{align}
\begin{split}
\big{[}\mathcal{O}_{lq}^{(1)}\big{]}_{ijkl} &= \big{(}\overline{L}_i\gamma^\mu L_j\big{)} \big{(}\overline{Q}_k \gamma_\mu Q_l\big{)}\,, \\[0.4em]
\big{[}\mathcal{O}_{lq}^{(3)}\big{]}_{ijkl} &=\big{(}\overline{L}_i\gamma^\mu \tau^ I L_j\big{)} \big{(}\overline{Q}_k \tau^ I\gamma_\mu Q_l\big{)}\,,\\[0.4em]
\big{[}\mathcal{O}_{eq}\big{]}_{ijkl} &=\big{(}\overline{e}_i\gamma^\mu  e_j\big{)} \big{(}\overline{Q}_k \gamma_\mu Q_l\big{)}\,,\\[0.4em]
\big{[}\mathcal{O}_{ld}\big{]}_{ijkl} &=\big{(}\overline{L}_i\gamma^\mu  L_j\big{)} \big{(}\overline{d}_k\gamma_\mu d_l\big{)}\,,\\[0.4em]
\big{[}\mathcal{O}_{ed}\big{]}_{ijkl} &=\big{(}\overline{e}_i\gamma^\mu  e_j\big{)} \big{(}\overline{d}_k\gamma_\mu d_l\big{)}\,,
\end{split}
\end{align}
where $ Q, L$ denote the SM quark and lepton $SU(2)_L$ doublets, and $u, d, e$ are the quark and lepton weak singlets. Flavor indices are denoted by $i,j,k,l$. In what follows, we work in the flavor basis defined with diagonal down-quark Yukawa matrix, i.e.~the CKM matrix appears in the upper component of $Q_i=[(V^\dagger\,u)_i\, ,\,d_i ]^T$.

The matching of the SMEFT Lagrangian to Eq.~\eqref{eq:eft-bsll} gives
\begin{align}
\label{eq:bsll-smeft}
\delta C_9^{\ell_i\ell_i}- \delta C_{10}^{\ell_i\ell_i} &= \dfrac{2\pi}{\alpha_\mathrm{em} \lambda_t} \dfrac{v^2}{\Lambda^2} \left\lbrace\big{[}\mathcal{C}_{lq}^{(1)}\big{]}_{ii23}+\big{[}\mathcal{C}_{lq}^{(3)}\big{]}_{ii23}\right\rbrace\,, \nonumber\\[0.4em]
\delta C_9^{\ell_i\ell_i}+ \delta C_{10}^{\ell_i\ell_i} &= \dfrac{2\pi}{\alpha_\mathrm{em} \lambda_t} \dfrac{v^2}{\Lambda^2} \big{[}\mathcal{C}_{qe}\big{]}_{ii23} \,,\nonumber\\[0.4em]
\delta C_{9^\prime}^{\ell_i\ell_i}-\delta C_{10^\prime}^{\ell_i\ell_i} &=\dfrac{2\pi}{\alpha_\mathrm{em} \lambda_t} \dfrac{v^2}{\Lambda^2} \big{[}\mathcal{C}_{\ell d}\big{]}_{ii23} \,, \\[0.4em]
\delta C_{9^\prime}^{\ell_i\ell_i}+\delta C_{10^\prime}^{\ell_i\ell_i} &= \dfrac{2\pi}{\alpha_\mathrm{em} \lambda_t} \dfrac{v^2}{\Lambda^2} \big{[}\mathcal{C}_{ed}\big{]}_{ii23}\,.\nonumber
\end{align}
The (pseudo)scalar Wilson coefficients are not explicitly written since their contributions to $\mathcal{B}(B\to K^{(\ast)}ll)$ are suppressed by the light lepton masses. 

Similarly, for the $b\to s\nu\bar{\nu}$ effective Lagrangian defined in Eq.~\eqref{eq:eft-bsnunu} we get:
\begin{align}
\begin{split}
\label{eq:left-CL-bsnunu1}
\delta C_L^{\nu_i \nu_j} &= \dfrac{\pi}{\alpha_\mathrm{em} \lambda_t} \dfrac{v^2}{\Lambda^2} \left\lbrace\big{[}\mathcal{C}_{lq}^{(1)}\big{]}_{ij23}-\big{[}\mathcal{C}_{lq}^{(3)}\big{]}_{ij23}\right\rbrace \,, \\[0.4em]
\delta C_R^{\nu_i \nu_j} &= \dfrac{\pi}{\alpha_\mathrm{em} \lambda_t} \dfrac{v^2}{\Lambda^2} \big{[}\mathcal{C}_{\ell d}\big{]}_{ij23}\,.
\end{split}
\end{align}

The above relations, together with the general expression given in  Eq.~\eqref{eq:ratio-magic-numbers}, allow us to predict $\smash{\mathcal{R}_{K^{(\ast)}}^{(\nu/l)}}$ in any extension of the SM.

\begin{figure*}[t!]
\centering
\includegraphics[width=0.47\linewidth]{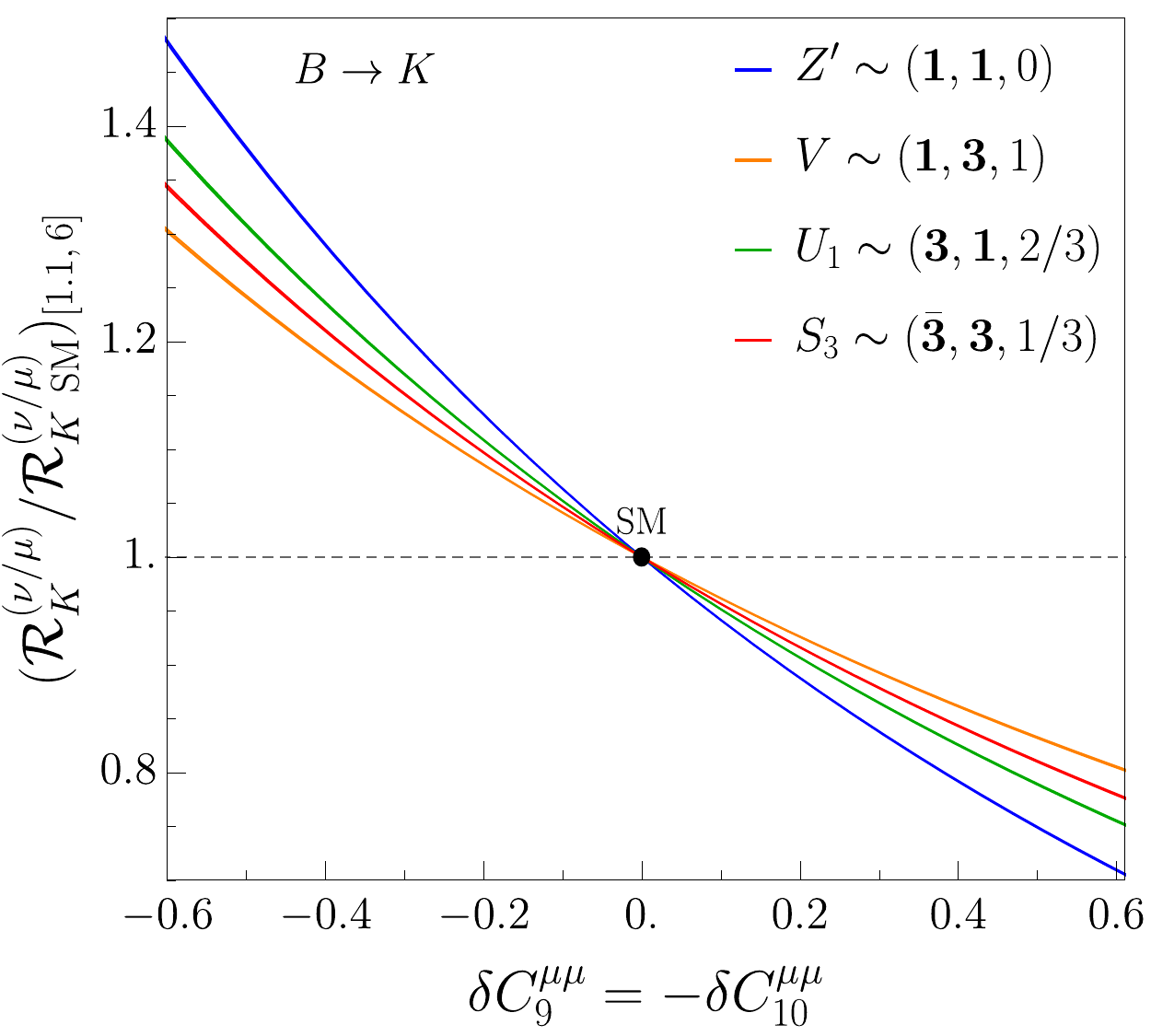}~\quad~\includegraphics[width=0.47\linewidth]{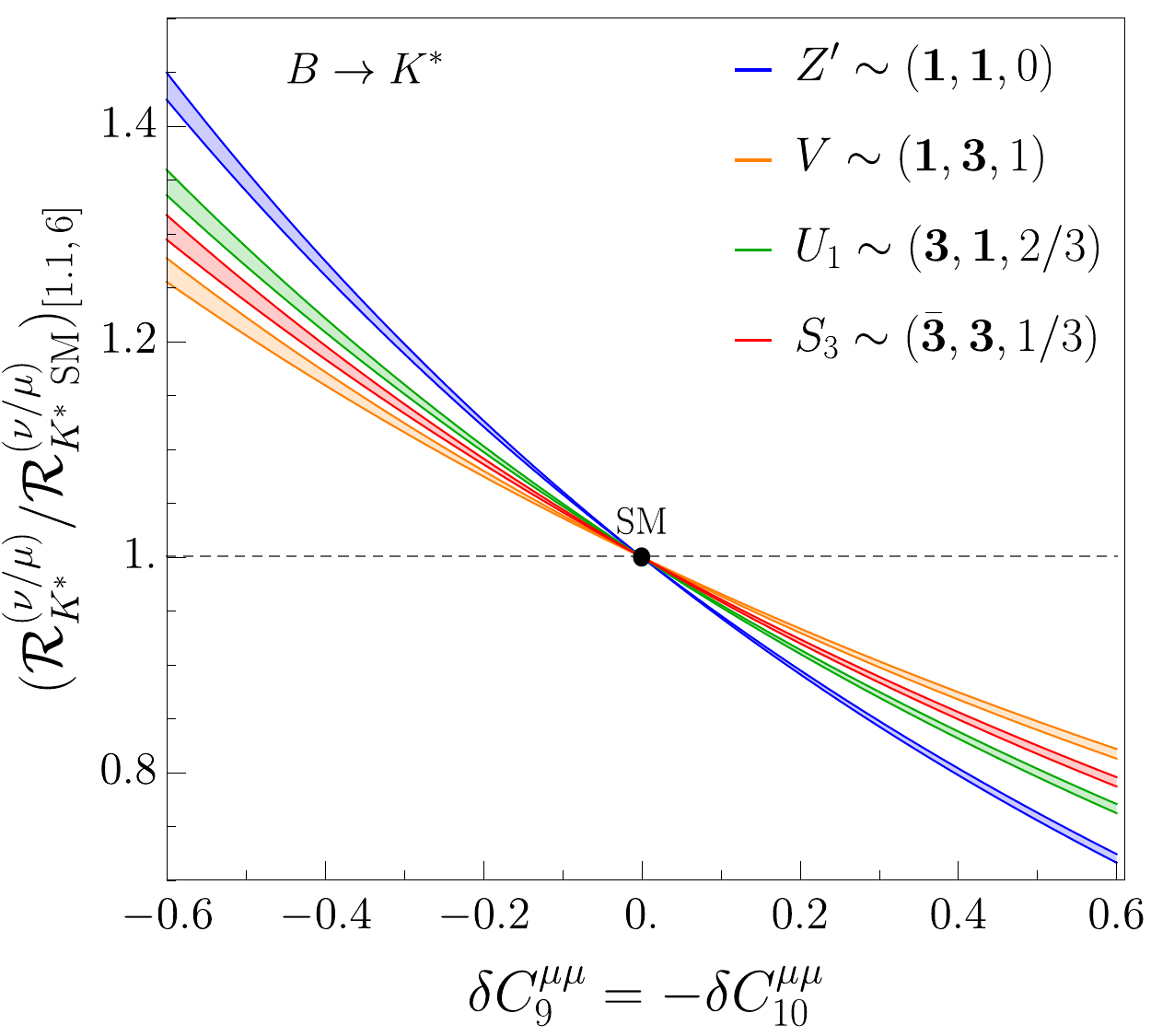}

\caption{\small \sl The ratios $\mathcal{R}_{K^{(\ast)}}^{(\nu/e)}$ (bottom and top left panels) and $\mathcal{R}_{K^{(\ast)}}^{(\nu/\mu)}$ (bottom and top right panels) normalized to their SM values, in the region $q^2\in [1.1,6]~\mathrm{GeV}^2$, and plotted against $\delta C_9^{\mu\mu}=-\delta C_{10}^{\mu\mu}$ for selected New Physics scenarios with left-handed couplings to muons, see Sec.~\ref{sec:bsm}. }
\label{fig:ratios-nu-l} 
\end{figure*}

\paragraph*{Illustration:} We will now consider the simplest scenario of New Physics that can contribute to the $b\to s\mu\mu$ decay modes, namely via left-handed effective operators satisfying $\delta C_9^{\mu\mu}=-\delta C_{10}^{\mu\mu}$, and illustrate its impact onto the ratios~\eqref{eq:nu-mu-ratio}. In terms of the SMEFT operators, this scenario can arise from any combination of $\smash{\mathcal{C}_{lq}^{(1)}}$ and $\smash{\mathcal{C}_{lq}^{(3)}}$, with couplings to muons, and to $2\to 3$ quark-flavor indices. Several concrete models can induce these operators through the exchange of new bosons \cite{Buras:2014fpa}. Clearly, these operators contribute not only to $b\to s\mu\mu$, but also to $b\to s\nu_\mu \bar{\nu}_\mu$, i.e.~to both the numerator and denominator of Eq.~\eqref{eq:nu-mu-ratio}. We classify them in terms of their SM quantum numbers $(SU(3)_c,SU(2)_L,U(1)_Y)$,
\begin{itemize}
    \item[$\bullet$] $Z^\prime \sim (\mathbf{1},\mathbf{1},0)$\,: One of the simplest scenarios is to extend the SM with a $Z^\prime$ that couples exclusively to left-handed quarks and leptons, 
        \begin{equation}
            \mathcal{L}_{Z^\prime} \supset \left[g_{Q}^{ij}\,\bar{Q}_i \gamma_\mu Q_j + g_{L}^{ij}\,\bar{L}_i \gamma_\mu L_j\right]\, Z^{\prime\,\mu} \,,
        \end{equation}
    where $g_Q$ and $g_L$ are Hermitian matrices. 
    In this case, we find that
        \begin{equation}
            \mathcal{C}_{lq}^{(1)}\neq 0\,,\qquad \mathcal{C}_{lq}^{(3)}=0\,.
        \end{equation}

    \item[$\bullet$] $V \sim (\mathbf{1},\mathbf{3},0)$\,: $Z^\prime$ could be a part of a weak triplet $V$ with the following interaction Lagrangian~\cite{Greljo:2015mma},
        \begin{equation}
            \qquad\mathcal{L}_{V} \supset \left[g_{Q}^{ij}\,\bar{Q}_i \tau^a\gamma_\mu Q_j + g_{L}^{ij}\,\bar{L}_i \tau^a\gamma_\mu L_j\right]\, V_a^{\mu}\,,
        \end{equation}
        
       \noindent where $g_Q$ and $g_L$ are again Hermitian matrices, and $\tau^a$ are the Pauli matrices, with $a=1,2,3$. In this case, 
        \begin{equation}
            \mathcal{C}_{lq}^{(1)}=0 \,,\qquad \mathcal{C}_{lq}^{(3)}\neq 0\,. 
        \end{equation}
    \item[$\bullet$] $S_3 \sim (\bar{\mathbf{3}},\mathbf{3},1/3)$\,: $S_3$ is the scalar leptoquark, often used in the literature~\cite{Buchmuller:1986zs,Hiller:2014yaa}. Its Yukawa interaction is given by
        \begin{equation}
            \mathcal{L}_{S_3} \supset y_L^{ij}\, \overline{Q_i^C} i\tau_2 \big{(}\vec{\tau}\cdot \vec{S}_3\big{)} L_j+\mathrm{h.c.}\,,
        \end{equation}
        
       \noindent which then imply at tree-level,
        \begin{equation}
            \mathcal{C}_{lq}^{(1)}=3\,\mathcal{C}_{lq}^{(3)}\,. 
        \end{equation}
    \item[$\bullet$] $U_1 \sim (\mathbf{3},\mathbf{1},2/3)$\,: Another leptoquark often used in literature is the vector leptoquark $U_1$ which interacts with the left-handed fermions via~\cite{Buchmuller:1986zs}, 
        \begin{equation}
            \mathcal{L}_{U_1} \supset x_L^{ij}\, \overline{Q_i} \gamma_\mu L_j\,U_1^\mu+\mathrm{h.c.}\,,
        \end{equation}
        
        \noindent from which we have,
        \begin{equation}
            \mathcal{C}_{lq}^{(1)}=\mathcal{C}_{lq}^{(3)}\,. 
        \end{equation}
        This combination of Wilson coefficient is peculiar as it contributes to $b\to s\ell\ell$ at tree-level, but not to $b\to s\nu\bar \nu$ due to a cancellation in Eq.~\eqref{eq:left-CL-bsnunu1}.
\end{itemize}

\noindent The concrete models listed above predict different patterns for the effective coefficients $\mathcal{C}_{lq}^{(1)}$ and $\mathcal{C}_{lq}^{(3)}$, which imply specific model-dependent correlations between the $b\to s \mu\mu$ and $b\to s \nu_\mu \nu_\mu$ transitions.

In Fig.~\eqref{fig:ratios-nu-l} we plot the New Physics contribution to the ratios $R^{(\nu/\mu)}_K$ (left panel) and $R^{(\nu/\mu)}_{K^\ast}$ (right panel) for each of the scenarios listed above. For illustration purposes we consider the $[1.1,6]~\mathrm{GeV}^2$ interval. The SMEFT operators induce not only a modification of the numerator, but also of the denominator, which actually enhances the effect up to $\mathcal{O}(40\%)$ in some models. In other words, besides the theoretical accuracy of the $\mathcal{R}_{K^{(\ast)}}^{(\nu/l)}$ ratios, which is improved compared to the separate branching fractions, these ratios also allow us to increase the sensitivity to the type of New Physics effects considered here.

The above discussion is based on a minimalistic assumption that only the muonic couplings in the $b\to s\ell\ell$ and $b\to s\nu\bar \nu$ transitions are affected by New Physics, but obviously that assumption can be changed. For example, one can test whether or not the New Physics couplings to $\tau$'s are significant~\cite{deGiorgi:2022vup,Descotes-Genon:2020buf}, such as the case in some leptoquark scenarios that can accommodate the LFU discrepancies in the $b\to c\tau\nu$ transition~\cite{Angelescu:2021lln}. Since these scenarios predict an enhancement of $\mathcal{B}(B\to K\nu_\tau \bar \nu_\tau)$~\cite{Becirevic:2022tsj,Fuentes-Martin:2020hvc,Gherardi:2020qhc}, it is clear that the effects depicted in Fig.~\ref{fig:ratios-nu-l} would only increase in this case.

\begin{table*}[!t]
\renewcommand{\arraystretch}{1.7}
\centering
\begin{tabular}{cccccccc}
\hline
$a_0^+$ & $a_1^+$ &  $a_2^+$ & $a_0^0$ & $a_1^0$ & $a_0^T$ & $a_1^T$ &  $a_2^T$\\ \hline\hline
$0.4742(62)$ & $-0.894(51)$ & $-0.44(14)$ & $0.2939(36)$ & $0.277(40)$ & $0.4752(92)$ & $-0.921(82)$ & $-0.53(35)$ \\ \hline
$1$ & $-0.2904$  & $-0.0347$  & $0.7480$  &  $0.1844$ &  $0.6558$ &  $-0.2193$ & $-0.0751$\\
$\cdot$ & $1$ & $0.7757$  & $0.2291$  &  $0.8527$ &  $-0.2569$ &  $0.5371$ & $0.2574$\\
$\cdot$ & $\cdot$ & $1$  & $0.1690$  & $0.8455$  & $-0.1029$  & $0.3700$  & $0.2653$\\
$\cdot$ & $\cdot$ & $\cdot$  & $1$ &  $0.4568$ &  $0.5232$ & $0.0314$  & $0.0257$ \\
$\cdot$ & $\cdot$ & $\cdot$  & $\cdot$ & $1$  &  $0.0182$ & $0.4501$  & $0.2372$\\
$\cdot$ & $\cdot$ & $\cdot$  & $\cdot$ & $\cdot$  & $1$  & $-0.0255$  & $-0.0535$\\
$\cdot$ & $\cdot$ & $\cdot$  & $\cdot$ & $\cdot$  & $\cdot$  & $1$  & $0.6920$\\
$\cdot$ & $\cdot$ & $\cdot$  & $\cdot$ & $\cdot$  & $\cdot$  & $\cdot$  & $1$\\
\hline
\end{tabular}
\caption{ \sl \small  \label{TabParams} Values of the $z$-expansion coefficients and correlation matrix obtained by our combined fit to the $B\to K$ form factors computed by HPQCD~\cite{Parrott:2022rgu} and FNAL/MILC~\cite{Bailey:2015dka} (with $\chi^2_\mathrm{min}/\mathrm{d.o.f}\simeq 9.2/10$). We consider the parameterization from Eq.~\eqref{eq:ff-par-fp}--\eqref{eq:ff-par-fT} with $N=3$, and we remove the $a_2^0$ coefficient by imposing the relation $f_+(0)=f_0(0)$. The covariance matrix is provided with more digits in an ancillary file. See text for more details. }
\label{tab:ff-fit}  
\end{table*}

\section{Summary}\label{sec:summary}

In this paper we revisited the SM estimate of $\mathcal{B}( B \to K^{(\ast )} \nu\bar \nu)$. This is particularly important for the case of the pseudoscalar meson in the final state because the relevant form factor has been extensively studied and computed by means of LQCD. Since the new such a calculation appeared after the most recent release of the FLAG review, we updated the FLAG average of all three form factors relevant to the $B\to K$ transitions. Since the lattice QCD results are obtained for large $q^2$'s one should be careful when addressing the issue of systematic uncertainties. For that reason we believe that the most reliable way to test the SM value for $\mathcal{B}^\prime( B \to K^{(\ast )} \nu\bar \nu)$ is in the region of large $q^2$'s. Furthermore, the experimental information on $r_\mathrm{lh}={\mathcal{B}(B \to K  \nu\bar{\nu})_{\mathrm{low}-q^2}/\mathcal{B}(B\to K \nu\bar{\nu})_{\mathrm{high}-q^2}}$ would be helpful to test the validity of the extrapolation of LQCD results (obtained at high-$q^2$) to low $q^2$'s. Besides the hadronic uncertainties in $\mathcal{B}( B \to K^{(\ast )} \nu\bar \nu)$, it is important to further improve the $\lambda_t = V_{tb} V_{ts}^\ast$ which, by virtue of the CKM unitarity, is related to the problem of reconciling the value of $V_{cb}$ extracted from exclusive and from inclusive semileptonic decays. 

Most of the uncertainties mentioned above actually cancel out if one considers the ratio of the partial decay rates of 
$\mathcal{B}^\prime( B \to K^{(\ast )} \nu\bar \nu)$ and of $\mathcal{B}( B \to K^{(\ast )} \ell\ell)$, which we denote by $\mathcal{R}_{K^{(\ast)}}^{(\nu/l)}$. The major hadronic uncertainty in both rates comes from the form factor which cancels out in $\mathcal{R}_{K^{(\ast)}}^{(\nu/l)}$. The uncertainty from the multiplicative CKM factor $\lambda_t$ cancels out as well. However, the price to pay is that the Wilson coefficient $C_9$, entering $\mathcal{B}( B \to K^{(\ast )} \ell\ell)$, becomes an obstacle because it is sensitive to the contribution from the non-local operator arising from the vector current couplings to $c\bar c$. In the literature that contribution is often estimated by using the quark-hadron duality or by resorting to the model calculations. If we stick to the SM, we show that from a measurement of the ratio $\mathcal{R}_{K^{(\ast)}}^{(\nu/l)}$ in a given interval of $q^2$'s (preferably below the first $c\bar c$ resonance), one can extract the value $\langle C_9^\mathrm{eff}\rangle$, and indeed check whether of not the sizable non-factorizable contribution would result in $\langle C_9^\mathrm{eff}\rangle_K \neq \langle C_9^\mathrm{eff}\rangle_{K^\ast}$, as sometimes argued in the literature. By $\langle C_9^\mathrm{eff}\rangle$ we denote the $C_9^\mathrm{eff}(q^2)$ averaged over the interval in which $\mathcal{R}_{K^{(\ast)}}^{(\nu/l)}$ is measured. 

To further support the benefits of measuring $\mathcal{R}_{K^{(\ast)}}^{(\nu/l)}$, we also illustrate how it can be used to look for the effects of physics BSM. In a scenario in which the New Physics contributed at low energy scales through left-handed couplings, both to quarks and to leptons, we find that  $\mathcal{R}_{K^{(\ast)}}^{(\nu/l)}$ would be more sensitive a test of presence of physics BSM than its numerator and/or its denominator separately. We also provided an illustration of such a scenario in several simple models. 


\section*{Acknowledgments}

We thank F.~Jaffredo and E.~Lunghi for discussions regarding the fits to the form factors computed in lattice QCD,  M.~Fedele and P.~Stangl for discussion regarding $C_9^\mathrm{eff}$, and F.~Mescia for useful exchanges. This project has received support from the European Union’s Horizon 2020 research and innovation programme under the Marie Skłodowska-Curie grant agreement No~860881-HIDDeN.

\appendix

\section{$B\to K^{(\ast)}$ form factors}\label{app:form factors}

\paragraph*{$B\to K$\,:} The definitions of the vector ($f_+$) and scalar ($f_0$) form factors are given in Eq.~\ref{eq:ff}, while the tensor form factor $f_T$ is defined as follows
\begin{align}
\hspace{-0.6em}\langle \bar{K}(k) | \bar{s}\sigma_{\mu\nu} b | \bar{B}(p) \rangle &= -i\left(p_\mu k_\nu - p_\nu k_\mu\right)  \dfrac{2 f_T(q^2)}{m_B+m_K}\,.
\end{align}
We consider the same parameterization for the $B\to K$ form factors as provided by FLAG~\cite{Aoki:2021kgd},
\begin{align}
\label{eq:ff-par-fp}
f_+ (q^2) &= \frac{1}{P_+(q^2)}\sum_{n=0}^{N-1}a_n^+\, \left[z^n-(-1)^{n-N}\frac{n}{N}z^N\right]\,,\\[0.4em]
\label{eq:ff-par-f0}
f_T (q^2) &= \frac{1}{P_T(q^2)}\sum_{n=0}^{N-1}a_n^T\, \left[z^n-(-1)^{n-N}\frac{n}{N}z^N\right]\,,\\[0.4em]
\label{eq:ff-par-fT}
f_0 (q^2) &= \frac{1}{P_0(q^2)}\sum_{n=0}^{N-1}a_n^0\, z^n\,,
\end{align}
where $a_n^i$ (with $i\in \lbrace 0,+,T \rbrace$) are numerical coefficients, the pole factors are given by
\begin{align}\label{eq:poleP}
P_i(q^2)= 1-q^2/M_i^2\,,
\end{align}
 with $M_+=M_T=5.4154\ {\mathrm{GeV}}$ and $M_0=5.711\ {\mathrm{GeV}}$ \cite{Lang:2015hza}, and $z\equiv z(q^2)$ reads
\begin{equation}
\label{eq:z}
z(q^2)=\frac{\sqrt{t_+ - q^2} - \sqrt{t_+ - t_0}}{\sqrt{t_+ - q^2} + \sqrt{t_+ - t_0}}\, ,
\end{equation}
where $t_+=(m_B + m_K)^2$ and we consider $t_0 = (m_B + m_K) (\sqrt{m_B}-\sqrt{m_K})^2$. Since the scalar and vector form factor satisfy $f_+(0)=f_0(0)$, it is possible to remove one of the coefficients in Eq.~\eqref{eq:ff-par-fp} and \eqref{eq:ff-par-f0}, which we take to be $a_2^0$,
\begin{equation}
\label{eq:a20}
a_2^0=\frac{ f_+(0) - a_0^0-a_1^0 \, z(q^2=0)}{z^2(q^2=0)}\,.
\end{equation}

\begin{figure}[t!]
\centering
\includegraphics[width=.98\linewidth]{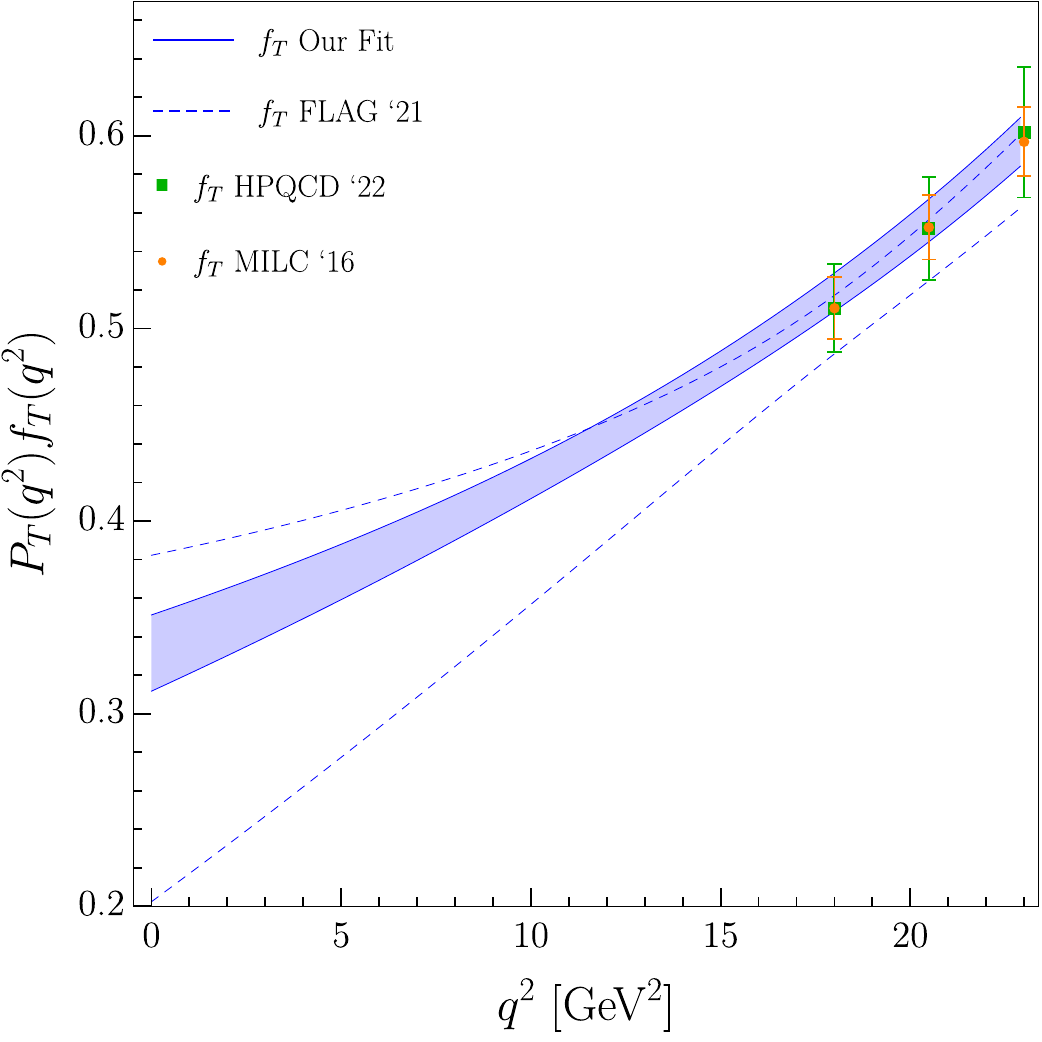}
\caption{\small \sl
The result of our fit for the  $f_{T}(q^2)$ form factors is depicted by the blue region respectively, compared to the result reported by FLAG~\cite{Aoki:2021kgd} and the HPQCD (green points)~\cite{Parrott:2022rgu} and FNAL/MILC~(orange points)~\cite{Bailey:2015dka} results. 
}
\label{fig:form factorFT} 
\end{figure}

In this letter, we update the combined fit to the $B\to K$ form factors made by FLAG~\cite{Aoki:2021kgd}, by including the latest HPQCD results~\cite{Parrott:2022rgu} that are combined with the ones from FNAL/MILC~\cite{Bailey:2015dka}. To this purpose, we follow the same procedure as FLAG which consists in generating synthetic data points for $\lbrace f_+(q^2),f_0(q^2),f_T(q^2)\rbrace$ for both HPQCD and FNAL/MILC form factors at three $q^2$ values, namely $q^2\in \lbrace 18, 20.5, 23\rbrace~\mathrm{GeV}^2$. The central values and the covariance matrix obtained for each collaboration are then fitted in a combined $\chi^2$ assuming that the HPQCD and FNAL/MILC results are uncorrelated. We consider the same parameterization of FLAG with $N=3$, as specified above, and we remove the $a_2^0$ coefficient by exploiting the scalar/vector form factor relations at $q^2=0$. 

The results of our $\chi^2$ fit are given in Table~\eqref{tab:ff-fit}, including the correlation matrix between form factor coefficients. Our combined fit gives $\chi^2_\mathrm{min}/\mathrm{d.o.f}\simeq 9.2/10$ and, differently from FLAG~\cite{Aoki:2021kgd}, we opt for not rescaling the uncertainties of the fitted parameters by $\sqrt{\chi^2_\mathrm{min}/\mathrm{d.o.f}}$. Our results for $f_+(q^2)$ are given in Fig.~\ref{fig:form factors}, where we see a good agreement between the two calculations and our combined fit.

We have performed several cross-checks of our fitting procedure. In particular, we are able to perfectly reproduce the FLAG results when combining FNAL/MILC~\cite{Bailey:2015dka} with the previous HPQCD results~\cite{Bouchard:2013eph}, provided we rescaled the uncertainties of the fitted parameters by $\smash{\sqrt{\chi^2_\mathrm{min}/\mathrm{d.o.f}}}$. As already stated above, we opt for not rescaling the uncertainties of our new combined fit.

\paragraph*{$B\to K^\ast$:}  The vector and axial form factors entering the $B\to K^\ast$ transition are $\{A_0$,  $A_1$, $A_{2}$, $V\}$, which are defined in Eq.~\ref{eq:BKst-ff}. The tensor form factors $\{T_1$, $T_2$, $T_{3}\}$ are defined by
\begin{align}
\label{eq:BKst-ff2}
\langle  \bar{K}^\ast(k) & | \bar{s}  \sigma_{\mu \nu}q^\nu(1-\gamma_5) b | \bar{B}(p) \rangle =
 2 i \varepsilon_{\mu\alpha\beta \gamma} \epsilon^{\ast \alpha} p^\beta k^\gamma T_1(q^2) \nonumber\\[0.9em]
+&  \bigl[(m_B^2 - m_{K^{\ast}}^2) \epsilon^{\ast}_{ \mu}-(\epsilon^\ast \cdot q) (k+p)_\mu \bigr]\,T_2(q^2) \nonumber\\
+&  (\epsilon^\ast \cdot q)\left[q_\mu -\dfrac{q^2}{m_B^2 - m_{K^{\ast}}^2}(k+p)_\mu \right]\,T_3(q^2) \,,
\end{align}

\noindent where $\varepsilon_\mu$ is the polarization vector of $K^\ast$. The $B\to K^\ast$ form factor parameterization and input parameters used in our study are taken from Ref.~\cite{Bharucha:2015bzk}, and we conservatively assume the fitted parameters to be uncorrelated.

\section{$B^+\to K^{(\ast)+}\nu\nu$ predictions at $\mathcal{O}(G_F^2)$}\label{app:shot-distance}

In Table~\ref{tab:binned-new}, we provide the predictions for the binned branching fraction of $B^+\to K^{(\ast)+}\nu\nu$ decays at $\mathcal{O}(G_F^2)$, i.e.~without including the tree-level contributions described in Eq.~\eqref{eq:treeDD} and \eqref{eq:treeDDsst}. These values are to be compared to the full predictions given in Tables~\ref{tab:binned-predictions-K} and~\ref{tab:binned-predictions-Kst}.

\begin{table*}[]
\renewcommand{\arraystretch}{1.9}
\centering
\begin{tabular}{|c||c|c||c|c|}
\hline 
$q^2$-bin $[\mathrm{GeV}^2]$ &  $\mathcal{B}(B^+ \to K^+ \nu\bar{\nu})_\mathrm{loop}\times 10^{6}$ & $\sigma_{\mathcal{B}_{K^+}}/\mathcal{B}_{K^+}$&  $\mathcal{B}(B^+ \to K^{\ast +} \nu\bar{\nu})_\mathrm{loop}\times 10^{6}$ & $\sigma_{\mathcal{B}_{K^{\ast+}}}/\mathcal{B}_{K^{\ast+}}$ \\ \hline\hline
$[0,4]$ & $(1.056 \pm 0.055\pm 0.064 )$ & $0.08$ & 
$({1.48 \pm 0.20 \pm0.09} )$ & $0.15$ \\ 
$[4,8]$ & $(1.028 \pm 0.039 \pm 0.062)$ & $0.07$ &
$( 2.00\pm  0.23\pm 0.12 )$ & $0.13$  \\ 
$[8,12]$ & $( 0.948\pm 0.027\pm 0.058)$ & $0.07$ &
$( 2.41 \pm0.30\pm0.15)$ & $0.14$  \\ 
$[12,16]$ & $( 0.790\pm 0.020\pm 0.048)$ & $0.07$ & 
$( 2.51\pm0.39\pm0.15)$ & $0.17$  \\ 
$[16,q^2_\mathrm{max}]$ & $( 0.614\pm 0.017\pm 0.037 )$ & $0.07$ &
$( 1.40\pm0.30\pm0.09 )$ & $0.23$ \\[0.3em] \cdashline{1-5}
$[0,q^2_\mathrm{max}]$ & $(4.44\pm 0.14\pm 0.27 )$ & $0.07$ &
$( 9.79\pm1.30 \pm 0.60)$ &  $0.15$  \\ \hline

\end{tabular}
\caption{ \sl \small SM predictions for the partially integrated $B^+\to K^{(\ast)+} \nu\bar{\nu}$ branching fractions, in a given $q^2$-bin, at $\mathcal{O}(G_F^2)$, i.e.~without including the tree-level annihilation contributions to these decays. The first uncertainty comes from the hadronic form factors and the second one is dominated by the uncertainty on $|\lambda_t|^2$. The total relative uncertainty of each observable is shown in the last column. See the captions of Tables~\ref{tab:binned-predictions-K} and~\ref{tab:binned-predictions-Kst} for the inputs considered.
}
\label{tab:binned-new} 
\end{table*}




\end{document}